\documentclass[journal=jacsat,manuscript=article]{achemso}
\usepackage{siunitx}
\usepackage{graphicx} 
\usepackage[utf8]{inputenc}
\usepackage[T1]{fontenc} 
\usepackage{amsmath}
\usepackage{amsfonts}
\usepackage{amssymb}
\usepackage{array}

\usepackage[version=3]{mhchem} 
\usepackage{braket}
\usepackage{bm}
\usepackage{xr}
\usepackage{xspace}
\usepackage{multirow}
\usepackage{titlesec}

\usepackage{xspace} 
\usepackage{caption}
\usepackage{subcaption}
\usepackage{tikz}
\usetikzlibrary{arrows, arrows.meta}
\usepackage{comment}
\usepackage{newtxtext}
\usepackage{mathtools}
\setkeys{acs}{articletitle = true}

\DeclareUnicodeCharacter{0301}{\'{e}}

\usepackage{xr-hyper}
\usepackage{hyperref}
\usepackage{cleveref}
\makeatletter
\newcommand*{\addFileDependency}[1]{
 \typeout{(#1)}
 \@addtofilelist{#1}
 \IfFileExists{#1}{}{\typeout{No file #1.}}
}
\makeatother

\newcommand*{\myexternaldocument}[1]{
 \externaldocument{#1}
 \addFileDependency{#1.tex}
 \addFileDependency{#1.aux}
}

\myexternaldocument{SI}

\newcommand*{\sm}{SI\xspace}

\author{Matteo Rinaldi}
\affiliation{Scuola Normale Superiore, Piazza dei Cavalieri 7, I-56126 Pisa, Italy}
\author{Chiara Sepali}
\affiliation{Scuola Normale Superiore, Piazza dei Cavalieri 7, I-56126 Pisa, Italy}
\author{Alicia Marie Kirk}
\affiliation{Scuola Normale Superiore, Piazza dei Cavalieri 7, I-56126 Pisa, Italy}
\author{Claudio Amovilli}
\affiliation{Dipartimento di Chimica e Chimica
Industriale, Universit\`a di Pisa, via Moruzzi 13, I-56124 Pisa, Italy}
\author{Chiara Cappelli}
\affiliation{Scuola Normale Superiore, Piazza dei Cavalieri 7, I-56126 Pisa, Italy}
\email{chiara.cappelli@sns.it}

\title{DMRG/FQ: a Polarizable Embedding Approach Combining Density Matrix Renormalization Group and Fluctuating Charges}
\abbreviations{IR,NMR,UV}

\begin{document}

\begin{tocentry}
\includegraphics[]{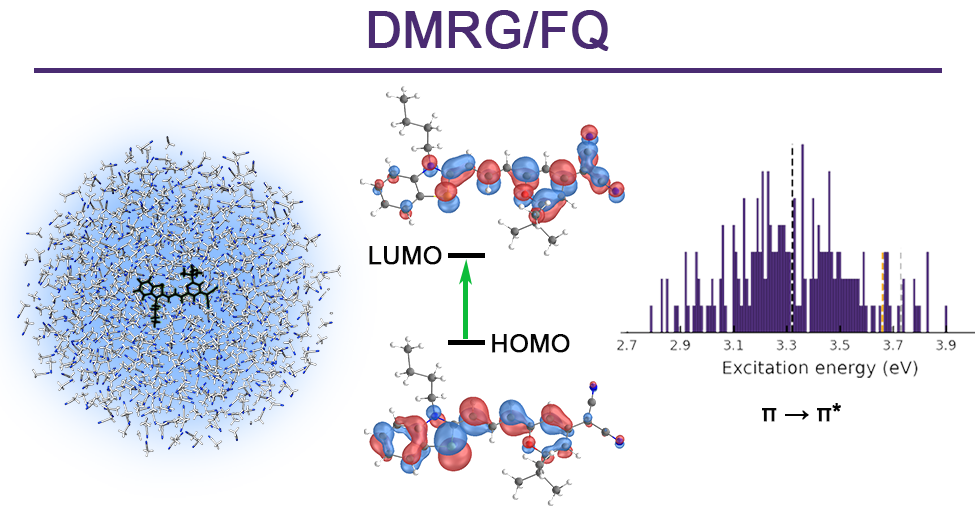}




\end{tocentry}

\begin{abstract}
We present an integrated multiscale framework that combines the Density Matrix Renormalization Group (DMRG) with a polarizable fluctuating-charge (FQ) force field for the simulation of electronic excited states in solution. The method exploits the capabilities of DMRG to accurately describe systems with strong static correlation, while the FQ model provides a self-consistent and physically grounded representation of solvent polarization within a QM/MM embedding.
The DMRG/FQ approach is applied to representative solvated systems, using extensive molecular dynamics sampling. The method yields reliable excitation energies, solvatochromic shifts, and a close agreement with available experimental data. The results highlight the importance of mutual polarization for capturing specific solute–solvent interactions, particularly in systems where hydrogen bonding or directional interactions play a dominant role.
\end{abstract}

\newpage

\section{Introduction}

Accurate modeling of electronically excited states in complex environments remains a central challenge in theoretical and computational chemistry.\cite{gomez2022multiple,Olsen2010PE,List2016PE,Mennucci2007CSM,Sneskov2011PECC,Neugebauer2007SubsystemDFT,Curutchet2017LightHarvesting,Humeniuk2024MSPE,hagras2018polarizable,giovannini2019dyes} In condensed-phase systems, solvent polarization and specific solute–solvent interactions can significantly alter the electronic structure of a chromophore,\cite{Zuehlsdorff2016SolventEffects,Beerepoot2014ConvergencePolarization,giovannini2020molecular,Masunov2005Paracyclophane,Dereka2017SoluteSolventSB,Gordon2001EFP} thereby affecting spectroscopic signatures, photochemical reactivity, and charge-transfer processes. Capturing these effects requires multiscale approaches capable of simultaneously treating electron correlation in the quantum region and the dynamic response of the surrounding environment.\cite{cas_fq,Humeniuk2024MSPE,hagras2018polarizable,cas_sn2_pcm,cas_amoeba,cas_mm_pol_nottoli}

A widely adopted strategy for incorporating environmental effects consists of using continuum embedding models, such as the Polarizable Continuum Model (PCM) and its variants \cite{qm_pcm,giovanninichemcomm}. These approaches describe the solvent as a polarizable dielectric medium defined by macroscopic parameters, providing an efficient and physically motivated route to account for bulk electrostatic polarization. PCM-based models have been successfully applied to many excitation phenomena, including the description of vertical excitation energies and solvatochromic shifts.\cite{Mennucci2006Solvatochromism,Cammi2000FastPCM,Improta2006SSPCM,Mennucci2012PCMExcitedStates,song_casscf_sa_pcm} However, their intrinsic nature prevents them from capturing specific, localized solute–solvent interactions, such as hydrogen bonding, $\pi$--$\pi$
stacking, or structural rearrangements within the first solvation shell. As a consequence, continuum treatments can underestimate environmental contributions when short-range interactions play a dominant role. \cite{giovanninichemcomm} 

To address these limitations, quantum mechanics/molecular mechanics (QM/MM) embedding schemes provide a more detailed representation of the environment by treating the solute at a QM level and the solvent explicitly at a classical MM level \cite{WARSHEL1976}. In their simplest fixed-charge formulation, QM/MM models can already capture structural and energetic features arising from specific solute–solvent contacts.\cite{FieldBashKarplus1990QM_MM,SennThiel2009QM_MMReview,LinTruhlar2007QM_MM,AcevedoJorgensen2010QM_MM} Their accuracy is substantially enhanced when polarizable MM models are employed,\cite{cappelli2016integrated,Olsen2010PE,thole1981molecular,Glover2012PolarizableQM_MM,giovannini2019fqfmu} as they allow the MM environment to respond to the QM electron density. This mutual polarization is essential for correctly describing directional interactions such as hydrogen bonds, charge–dipole couplings, and the stabilization of charge-transfer excited states.\cite{giovannini2019irfqfmu} Among the various formulations, the coupling of QM wavefunctions with the Fluctuating Charge (FQ) force field\cite{rick1994dynamical,lipparini2011fq}  has emerged as particularly attractive due to its physical grounding in charge equilibration principles and computational efficiency.\cite{cappelli2016integrated,giovannini2020molecular,cas_fq}

The Density Matrix Renormalization Group (DMRG) offers a robust wavefunction-based method for treating systems that exhibit pronounced static correlation, especially when large active spaces are required.\cite{dmrg_baiardi_reiher,chan_head_gordon,chan_2016,wouters2014,molmps_veis,Yanai_2011}  Its tensor-network formulation in terms of matrix product states (MPSs) and matrix product operators (MPOs) \cite{SCHOLLWOCK}, together with orbital optimization, enables a flexible and accurate representation of the multiconfigurational electronic wavefunction \cite{Yanai_Chan,sharma_chan,dmrg_baiardi_reiher}. 

Extending DMRG to solvated systems, therefore, necessitates embedding schemes capable of handling both the long-range solvent response and localized interactions at the QM/MM boundary. Despite the individual successes of DMRG and polarizable QM/MM techniques, their integration for the simulation of excited states in solution has remained limited. 

One possible strategy for incorporating environmental polarization effects in DMRG consists of a fully quantum-mechanical treatment, in which the environment surrounding the DMRG subsystem is kept frozen and represented by an effective embedding potential \cite{fde_dmrg}. This approach is known as Frozen Density Embedding (FDE) \cite{FDE_1993} and environmental polarization is taken into account through iterative freeze-and-thaw cycles \cite{WESOLOWSKI_1996}. WFT-in-WFT\cite{multi_in_dft} embedding strategies based on a DMRG wave function have also been proposed, most notably within the framework of Density Matrix Embedding Theory (DMET) \cite{dmet_chan_1,dmet_chan_2,fde_chan_wouters}.

Environmental polarization effects can also be incorporated by employing DMRG to describe the quantum-mechanical region within a QM/MM framework \cite{dmrg_pol_hedegard}. To the best of our knowledge, the only previous attempt to introduce MM polarization in DMRG has resorted to a polarizable embedding based on induced dipoles. \cite{induced_dipole_theory}. The resulting calculations were performed in a DMRG-CI framework (i.e., not within a self-consistent field scheme), while dynamical correlation effects were recovered through a short-range DFT correction using the DMRG-srDFT ansatz \cite{dmrg_srDFT}.

In this work, we propose an integrated DMRG/FQ multiscale methodology, and we specialize it to the calculation of electronic excitation energies in solution. The method couples a fully optimized DMRG wavefunction with a polarizable FQ environment within a QM/MM formalism, allowing for mutual electrostatic polarization between the two subsystems. 

The paper is organized as follows: the next section briefly recalls the fundamentals of DMRG in the MPS-MPO formulation and the FQ force field. Then, the DMRG/FQ coupling is discussed. After a section explaining the computational protocols, the performance and capabilities of DMRG/FQ approach are illustrated on the calculation of excitation energies of representative solvated systems, including acetone in aqueous solution and a merocyanine dye (DCBT, see below) in acetonitrile, using extensive molecular dynamics sampling to characterize the distribution of excitation energies. A brief section summarizing the main results of this study ends the presentation.

\section{Theory}
In this section, the DMRG method in the MPS-MPO formulation and the super-CI approach, used for orbital optimization, are briefly recalled. Then, after a brief presentation of the fluctuating charge (FQ) force field, the coupling between DMRG and FQ is discussed.

\subsection{The DMRG method in the MPS-MPO formulation}\label{sec:dmrg}
The DMRG method was originally developed for the study of one-dimensional lattice systems \cite{White_1992,White_1993}. Its first formulation, based on renormalized blocks, was later introduced into quantum chemistry \cite{dmrg_qc_white_martin,ortolani,chan_head_gordon,sharma_chan}. More recently, the modern formulation relying on matrix product states (MPSs) and matrix product operators (MPOs) has been widely adopted \cite{ostlund_rommer,rommer_ostlund,SCHOLLWOCK,keller,chan_2016,dmrg_baiardi_reiher}.

The derivation of DMRG starts from the Complete Active Space Self Consistent Field (CASSCF) wavefunction, which, for an active space of $L$ orbitals, can be written as: \cite{roos1980complete,roos1980complete1}

\begin{equation}\label{eq:fcisq}
\ket{\Psi}= \sum_{\sigma_1,...,\sigma_L}\textbf{C}_{\sigma_1...\sigma_L}\ket{\sigma_1...\sigma_L}
\end{equation}

\noindent 
where

$\sigma_i$ denotes the occupation number of the $i$-th orbital, and $\mathbf{C}_{\sigma_1...\sigma_L}$ is the CASSCF coefficient tensor. The number of parameters in this wavefunction scales as $4^L$. 
The corresponding molecular Hamiltonian operator is: 
\begin{equation}\label{eq:ham}
  \hat{H} =  \sum_{pq} h_{pq} \hat{E}_{pq} + \frac{1}{2} \sum_{pqrs} g_{pqrs} \left( \hat{E}_{pq} \hat{E}_{rs} - \delta_{rq} \hat{E}_{ps} \right)
\end{equation}

\noindent
where $h_{pq}$ and $g_{pqrs}$ denote the one- and two-electron integrals in the molecular-orbital basis and $\hat{E}_{pq}$ is  the singlet excitation operator that acts on the molecular orbitals $p$ and $q$. 

In the modern formulation of DMRG, \cite{dmrg_baiardi_reiher} the CASSCF wavefunction is expressed as an MPS, by performing $L$ successive singular value decompositions (SVDs) of $\mathbf{C}_{\sigma_1...\sigma_L}$, yielding:

\begin{equation}\label{eq:mpform}
\ket{\Psi}= \sum_{\sigma_1,...,\sigma_L}\textbf{M}^{\sigma_1} \textbf{M}^{\sigma_2}\cdot\cdot\cdot \textbf{M}^{\sigma_L}\ket{\sigma_1...\sigma_L}
\end{equation}

\noindent
where the dimension of each matrix $\textbf{M}^{\sigma_i}$ resulting from the SVD is truncated to $M$, referred to as the maximum bond dimension.\cite{SCHOLLWOCK}
This truncation reduces the number of wavefunction parameters to $4 L M^2$, thus lowering the scaling from exponential to polynomial.

\noindent
This approach also requires representing operators in matrix product form \cite{SCHOLLWOCK}. In the MPO formalism, the Hamiltonian in eq. \ref{eq:ham} becomes:
   \begin{equation}
   \hat{H}=\sum_{b_1,...,b_{L-1}} H^1_{1b_1} \cdot\cdot\cdot H^l_{b_{l-1}b_l}\cdot\cdot\cdot H^L_{b_{L-1}1}
\end{equation}

\noindent
The expectation value of $\hat{H}$ is then:
\begin{align}
&\bra{\Psi}\hat{H}\ket{\Psi} =
\sum_{\substack{\sigma_L, \sigma_L'\\a_{L-1},a'_{L-1},b_{L-1}}} M_{1a_{L-1}}^{\sigma_L \dagger}H_{b_{L-1}1}^{\sigma_L \sigma '_L} \bigg{(} \cdot\cdot\cdot \\ & \sum_{\substack{\sigma_2, \sigma_2'\\a_1,a'_1,b_1}}M_{a_2a_1}^{\sigma_2 \dagger}H_{b_1b_2}^{\sigma_2 \sigma '_2} \big{(}\sum_{\sigma_1,\sigma_1'}M_{a_11}^{\sigma_1 \dagger}H_{1b_1}^{\sigma_1 \sigma '_1}M_{1a'_1}^{\sigma_1'}\big{)}M_{a'_1a'_2}^{\sigma_2'}\cdot\cdot\cdot\bigg{)}M_{a'_{L-1}1}^{\sigma_L'} \notag
\end{align}

\noindent
This expression can be simplified by defining the so called left boundaries ($\textbf{L}$) and right boundaries ($\textbf{R}$) as follows:\cite{SCHOLLWOCK, keller}
\begin{equation} \label{eq:leftboundary}
    \textbf{L}_{a_la'_l}^{b_l} = \sum_{\substack{\sigma_l, \sigma_l'\\a_{l-1},a'_{l-1},b_{l-1}}}M_{a_la_{l-1}}^{\sigma_l \dagger}H_{b_{l-1}b_l}^{\sigma_l \sigma '_l}  \textbf{L}_{a_{l-1}a'_{l-1}}^{b_{l-1}} M_{a'_{l-1}a'_l}^{\sigma_l'} 
\end{equation}

\begin{equation}  \label{eq:rightboundary}
    \textbf{R}_{a'_{l-1}a_{l-1}}^{b_{l-1}} = \sum_{\substack{\sigma_l, \sigma_l'\\a_l,a'_l,b_{l-1}}}M_{a'_{l-1}a'_l}^{\sigma_l'}H_{b_{l-1}b_l}^{\sigma_l \sigma '_l}  \textbf{R}_{a'_la_l}^{b_l} M_{a_la_{l-1}}^{\sigma_l' \dagger} 
\end{equation}

\noindent
As a result, the Hamiltonian expectation value takes the form:
\begin{align}
   &\bra{\Psi}\hat{H}\ket{\Psi}=\sum_{a_l,a'_l,b_l} \textbf{L}_{a_la'_l}^{b_l}\textbf{R}_{a'_la_l}^{b_l} = \\
   &\sum_{\substack{\sigma_l, \sigma_l'\\a_{l-1},a'_{l-1},a_l,a'_l\\b_{l-1},b_l}}M_{a_la_{l-1}}^{\sigma_l \dagger}H_{b_{l-1}b_l}^{\sigma_l \sigma '_l}  \textbf{L}_{a_{l-1}a'_{l-1}}^{b_{l-1}} M_{a'_{l-1}a'_l}^{\sigma_l'}\textbf{R}_{a'_la_l}^{b_l} \notag
\end{align}

\noindent
To variationally minimize the energy, a constrained minimization that preserves the normalization of the wavefunction must be performed. This is achieved by taking the derivative with respect to each tensor $M^{\sigma_l*}_{a_{l-1}a_l}$, leading to the following generalized eigenvalue problem: \cite{SCHOLLWOCK, keller}

\begin{equation}\label{eq:dmrg_energy}
  \sum_{\substack{\sigma_l, \sigma_l'\\a'_{l-1},a'_l,b_{l-1},b_l}}H_{b_{l-1}b_l}^{\sigma_l \sigma '_l}  \textbf{L}_{a_{l-1}a'_{l-1}}^{b_{l-1}} M_{a'_{l-1}a'_l}^{\sigma_l'}\textbf{R}_{a'_la_l}^{b_l}= E_{DMRG} \; M^{\sigma_l}_{a_{l-1}a_l}
\end{equation}

\noindent
which is solved using sparse eigensolver techniques, such as the Jacobi–Davidson algorithm. The procedure is performed for each tensor (or for each pair, if a two-site algorithm is employed\cite{keller,dmrg_baiardi_reiher}), moving back and forth in a process called a sweep, until convergence is reached.\cite{SCHOLLWOCK}

In CASSCF calculations performed with DMRG, the MPS optimization replaces the calculation of CI coefficients, while the orbital optimization is performed by resorting to specific techniques such as the super-CI approach \cite{roos1980complete,roos1980complete1}. 

In this framework, the orbital optimization is achieved by satisfying the condition:
\begin{align}\label{eq:cas_orb_gradient}
    g_{rs}^{(o)} &= \braket{\Psi| [\hat{H}, \hat{E}_{rs}^-]|\Psi} = 2 \braket{\Psi| \hat{H} \hat{E}_{rs}^-|\Psi} = 2 \braket{\Psi|\hat{H}|rs} = 0
\end{align}

\noindent
where $\hat{E}_{rs}^- = \hat{E}_{rs} -\hat{E}_{sr}$ and $\ket{rs} = \hat{E}_{rs}^-\ket{\Psi}$ are the so-called Brillouin states \cite{roos1980complete}. Eq. \ref{eq:cas_orb_gradient} is the result of  the Brillouin-Levy-Berthier (BLB) theorem, also known as the \textit{Extended Brillouin Theorem}. 
\cite{roos1992multiconfigurational,levy1968generalized,levy1969generalized} 
In case of DMRGSCF calculations, the sweep process described above is alternated with the super-CI procedure until convergence is achieved. This is the methodology implemented in Openmolcas \cite{aquilante2020modern, li2023openmolcas}, where the DMRG solver is called from QCMaquis\cite{keller,QCMaquis_v_4}, to which Openmolcas is interfaced. 
 
\subsection{The Fluctuating Charges (FQ) force field}\label{sec:fq}


The FQ polarizable force field\cite{rick1994dynamical,rick1995fluctuating}, describes each atom in the classical portion of the system in terms of a charge $q_{i \alpha}$ that is not fixed (such as in most classical force-fields) but "fluctuates" in response to the presence of the other portions of the system. The total FQ energy functional is given by a second-order Taylor expansion of the energy with respect to charges:

\begin{equation}\label{eq:fq_lagrangian}
\begin{aligned}
E_{\text{FQ
}} & = \sum_{i \alpha} \text{q}_{i \alpha} \chi_{i \alpha} + \frac{1}{2} \sum_{i \alpha} \sum_{j \beta} \text{q}_{i \alpha} \text{T}_{i \alpha, j \beta}^{\text{qq}} \text{q}_{j \beta} + \sum_{\alpha} \left[ \lambda_\alpha \sum_i (\text{q}_{i \alpha}) - \text{Q}_{\alpha} \right]
\end{aligned}
\end{equation}

\noindent
In this expression, the ($i, j$) and ($\alpha, \beta$) indices run over FQ atoms and molecules, respectively. $\chi_{i\alpha}$ indicates the atomic electronegativity, while 
$\text{T}_{i\alpha,j\beta}^{\text{qq}}$ is the charge-charge interaction kernel, whose diagonal elements $\text{T}_{i\alpha,i\alpha}^{\text{qq}}$ are defined from the chemical hardness $\eta_{i\alpha}$. FQ specifically employs the Ohno kernel\cite{ohno1964some}, to avoid the so-called "polarization catastrophe". The set of Lagrangian multipliers $\lambda_\alpha$ is introduced to constrain the total charge of each FQ moiety to $Q_\alpha$, thus preventing unphysical charge transfer effects. Note that FQ depends only on two parameters, $\chi_{i\alpha}$ and $\eta_{i\alpha}$, which can be rigorously defined in the framework of Conceptual Density Functional Theory.\cite{chelli2002transferable,mortier1985electronegativity} 

\noindent
 FQ atomic charges are obtained according to the Electronegativity Equalization Principle (EEP)\cite{sanderson1951interpretation}. In practice, they are computed by imposing stationarity conditions on the energy functional with respect to the atomic charges and the associated Lagrange multipliers, which leads to solving the following linear system:\cite{cappelli2016integrated}

\begin{equation}\label{eq:fq_syst}
\left(
\begin{array}{cc}
\mathbf{T}^{\text{qq}} & \mathbf{1}_{\lambda} \\ 
\mathbf{1}^{\dagger}_{\lambda} & \mathbf{0}  \\
\end{array}
\right)
\left({\begin{array}{c} 
\mathbf{q}\\
\lambda \\
\end{array}}\right)
=
\left(\begin{array}{c} - \chi \\ 
\mathbf{Q}_{\alpha} 
\end{array}\right) 
\end{equation} 

\noindent
where $\mathbf{1}_{\lambda}$ are rectangular blocks associated with Lagrange multipliers.

\subsection{The DMRG/FQ approach}\label{sec:dmrg-fq}


In line with previous studies of our group \cite{cappelli2016integrated,ambrosetti_parametrization,cas_fq}, the coupling between DMRG and the FQ force field is carried out within a quantum mechanics/molecular mechanics (QM/MM) framework. Accordingly, the total energy of a system described by the DMRG/FQ approach is given by:

\begin{equation}\label{eq:tot_energy}
    E = E_{\text{DMRG}} + E_{\text{FQ}} + E^\mathrm{int}_{\text{DMRG/FQ}}
\end{equation}

\noindent
where $E_{\text{DMRG}}$ is defined from eq. \ref{eq:dmrg_energy} and $E_{\text{FQ}}$ from eq. \ref{eq:fq_lagrangian}. In this paper, the interaction term in eq. \ref{eq:tot_energy} is formulated by limiting to the electrostatic interaction between the FQ charges and the quantum (DMRG) part, i.e.:

\begin{equation} \label{eq:elec_int_term}
    E^\mathrm{int}_{\text{DMRG/FQ}} = \sum_{i\alpha} q_{i\alpha} V_{i\alpha}(\mathbf{D})
\end{equation}

\noindent
where $\mathbf{D}$ is the QM one-particle density matrix, and $V_{i\alpha}(\mathbf{D})$ is the total electrostatic potential acting on the FQ charge $q_{i\alpha}$ at position $\mathbf{r}_{i\alpha}$. It is defined as:
\begin{equation} \label{eq:elec_pot}
V_{i\alpha}(\mathbf{D}) = \sum_{N}^\mathrm{nuclei} \frac{Z_N}{\left|\mathbf{r}_{i\alpha}-\mathbf{R}_N\right|} - \sum_{pq}D_{pq}V^\mathrm{FQ}_{pq,i\alpha},\quad V^\mathrm{FQ}_{pq,i\alpha} = \braket{\phi_p|\frac{1}{\left|\mathbf{r}-\mathbf{r}_{i\alpha}\right|}|\phi_q}
\end{equation}

\noindent
In eq. \ref{eq:elec_pot}, the first term is the potential generated by the nucleus $N$ with charge $Z_N$ located at the position $\mathbf{R}_N$. The second term is the electronic potential expressed in terms of $\mathbf{D}$.

Hence, from eq. \ref{eq:tot_energy}, the total DMRG/FQ energy functional becomes:
\begin{equation}
\begin{split}\label{eq:dmrg_fq_functional}
    E_\mathrm{DMRG/FQ} (\mathbf{D}, \mathbf{P},  \mathbf{q}, \bm{\lambda}) = & \; E_\mathrm{DMRG} (\mathbf{D}, \mathbf{P}) + \sum_{i\alpha} q_{i\alpha} \chi_{i\alpha} + \frac{1}{2} \sum_{i\alpha, j\beta} q_{i\alpha} T_{i\alpha, j\beta}^\mathrm{qq} q_{j\beta } \\ & + \sum_{i\alpha} q_{i\alpha} V_{i\alpha}(\mathbf{D}) + \sum_\alpha \lambda_\alpha \left[\sum_i q_{i\alpha } - Q_{\alpha}\right] 
\end{split}
\end{equation}

\noindent
where $\mathbf{P}$ represents the two-particle density matrix. In line with a previous study of some of us\cite{cas_fq}, a state-specific (SS) approach is used to define the densities, i.e. one-particle and two-particle density matrices come from a single selected state.
\noindent
The FQ charges of eq. \ref{eq:dmrg_fq_functional} are obtained by minimizing the  DMRG/FQ energy functional with respect to FQ charges and Lagrange multipliers $\lambda_\alpha$. In this way, a linear system like that of eq. \ref{eq:fq_syst} is obtained, which is modified by accounting for the QM potential as an additional polarization source:

\begin{equation}
\left(
\begin{array}{cc}
\mathbf{T}^{\text{qq}} & \mathbf{1}_{\lambda} \\ 
\mathbf{1}^{\dagger}_{\lambda} & \mathbf{0}  \\
\end{array}
\right)
\left({\begin{array}{c} 
\mathbf{q}\\
\bm{\lambda}\\ 
\end{array}}\right)
=
\left(\begin{array}{c} -\bm{\chi} \\ 
\mathbf{Q}_{\alpha} \\
\end{array}\right) + \left(\begin{array}{c} -\mathbf{V} (\mathbf{D})\\ 
\mathbf{0} \\
\end{array}\right)
\label{eq:lin_sys_qm_fq}
\end{equation}

\noindent
Since the interaction term in eq. \ref{eq:elec_int_term} is monoelectronic, it is inserted into the one-electron integrals of the molecular hamiltonian, resulting in the following effective Hamiltonian:

\begin{equation}\label{eq:ham_eff}
  \hat{H}^\mathrm{eff} =  \sum_{pq} \left[h_{pq} + \mathbf{q}^\dagger \mathbf{V}^\mathrm{FQ}_{pq}\right]  \hat{E}_{pq} + \frac{1}{2} \sum_{pqrs} g_{pqrs} \left( \hat{E}_{pq} \hat{E}_{rs} - \delta_{rq} \hat{E}_{ps} \right)
\end{equation}

\noindent
which will be expressed as an MPO. To minimize the energy, analogously to eq. \ref{eq:dmrg_energy}, the effective Hamiltonian in eq. \ref{eq:ham_eff}---with the inclusion of the explicit FQ contribution---is diagonalized by solving the following generalized eigenvalue problem, yielding the DMRG/FQ energy:

\begin{equation}\label{eq:dmrg_fq_energy}
  \sum_{\substack{\sigma_l, \sigma_l'\\a'_{l-1},a'_l,b_{l-1},b_l}}H_{b_{l-1}b_l}^{eff \; \sigma_l \sigma '_l}  \textbf{L}_{a_{l-1}a'_{l-1}}^{b_{l-1}} M_{a'_{l-1}a'_l}^{\sigma_l'}\textbf{R}_{a'_la_l}^{b_l}= E_{DMRG/FQ} \; M^{\sigma_l}_{a_{l-1}a_l}
\end{equation}

\noindent
 The eigenvalue problem in eq.~\ref{eq:dmrg_fq_energy} is solved alternately with orbital optimization and the calculations of the FQ charges from eq. \ref{eq:lin_sys_qm_fq}, until energy convergence is achieved. Orbital optimization is carried out by including the FQ contributions in the orbital gradient used in the super-CI procedure, in a manner analogous to the CASSCF/FQ approach described in our previous work \cite{cas_fq}. The expression for the orbital gradient including the FQ terms is given in eq.~\ref{eq:dmrg_fq_orb_gradient}, where the term \( g_{rs}^{(o)} \) corresponds to the expression given in eq.~\ref{eq:cas_orb_gradient}:

\begin{align}\label{eq:dmrg_fq_orb_gradient}
    g_{\mathrm{tot},rs}^{(o)} & = 
    g_{rs}^{(o)} + 2 \braket{\Psi_0| \sum_{pq} \mathbf{q}^\dagger\mathbf{V}^\mathrm{FQ}_{pq} \hat{E}_{pq}|rs}
\end{align}

\noindent

\noindent
In summary, a DMRGSCF/FQ calculation requires: 

\begin{enumerate}
    \item Computing starting orbitals;
    \item Optimizing the MPS and obtaining the initial density matrices, $\mathbf{D}^{(0)}$ and $\mathbf{P}^{(0)}$, through eq. \ref{eq:dmrg_energy};
    \item Computing the starting FQ charges $\mathbf{q}^{(0)}$ from eq. \ref{eq:lin_sys_qm_fq};
    \item for $k=1,2,\dots$ until convergence:
    \begin{enumerate}
        \item The MPS optimization and the density matrices $\mathbf{D}^{(k)}$, $\mathbf{P}^{(k)}$ are computed with the inclusion of FQ contributions through eq. \ref{eq:dmrg_fq_energy};
        \item The molecular orbitals $\mathbf{T}^{(k)}$ are optimized with the inclusion of FQ contributions in eq. \ref{eq:dmrg_fq_orb_gradient};
        \item The FQ charges $\mathbf{q}^{(k)}$ are updated from eq. \ref{eq:lin_sys_qm_fq};
        \item The SS-DMRGSCF/FQ energy is finally computed by means of eq. \ref{eq:dmrg_fq_functional}.
    \end{enumerate}
\end{enumerate}

For brevity, hereafter we denote DMRGSCF/FQ as DMRG/FQ.

\section{Computational details}\label{sec:comp_det}

In this work, the vertical excitation energies of acetone in aqueous solution and of the merocyanine dye 4-(dicyanomethylene)-2-tert-butyl-6-[3-(3-butyl-benzothiazol-2-ylidene)-1-propenyl]-4H-pyran (DCBT)\cite{paper_DCBT} in acetonitrile were computed to assess the quality of the proposed approach.  
A multi-step protocol---adapted from previously established methodology specifically developed for modeling spectral signals of solvated molecules at the QM/MM level~\cite{giovannini2020molecular}---was employed as follows:  

\begin{enumerate}
\item \textit{Definition of the system}: The solutes (acetone and DCBT) were treated at the QM (DMRG) level, while the solvents (water and acetonitrile) were described at the MM level, using the polarizable FQ force field.

\item \textit{Conformational Sampling}: An accurate sampling of the possible solute--solvent configurations in solution was obtained by performing classical, non-polarizable molecular dynamics (MD) simulations over a timescale of tens of nanoseconds. 
For acetone, a previous 20 ns MD simulation of acetone in water (TIP3P) was utilised (which employed customised parameters for acetone - 
$MD _{REFINED}$)\cite{skoko}. For DCBT, a 30 ns MD simulation of DCBT in acetonitrile (NVT) was performed with the GROMACS package \cite{abraham2015gromacs} using the general AMBER force ﬁeld (GAFF)\cite{gaff} and acetonitrile parameters from Kowsari and coworkers\cite{kowsari}  [see Section S1 in the Supporting Information (\sm) for further details].    

\item \textit{Extraction of Structures}: 
A set of uncorrelated snapshots were extracted from the production phase of the MD simulations of acetone and DCBT. For each configuration, a solute-centered spherical droplet was generated using radii of 15 \AA{} for acetone and 30 \AA{} for DCBT to retain the relevant solute–solvent interactions. Example configurations for both systems are illustrated in \cref{f:plot_acetone_and_DCBT}.
\begin{figure}[ht!]
 \includegraphics[scale=1.0]{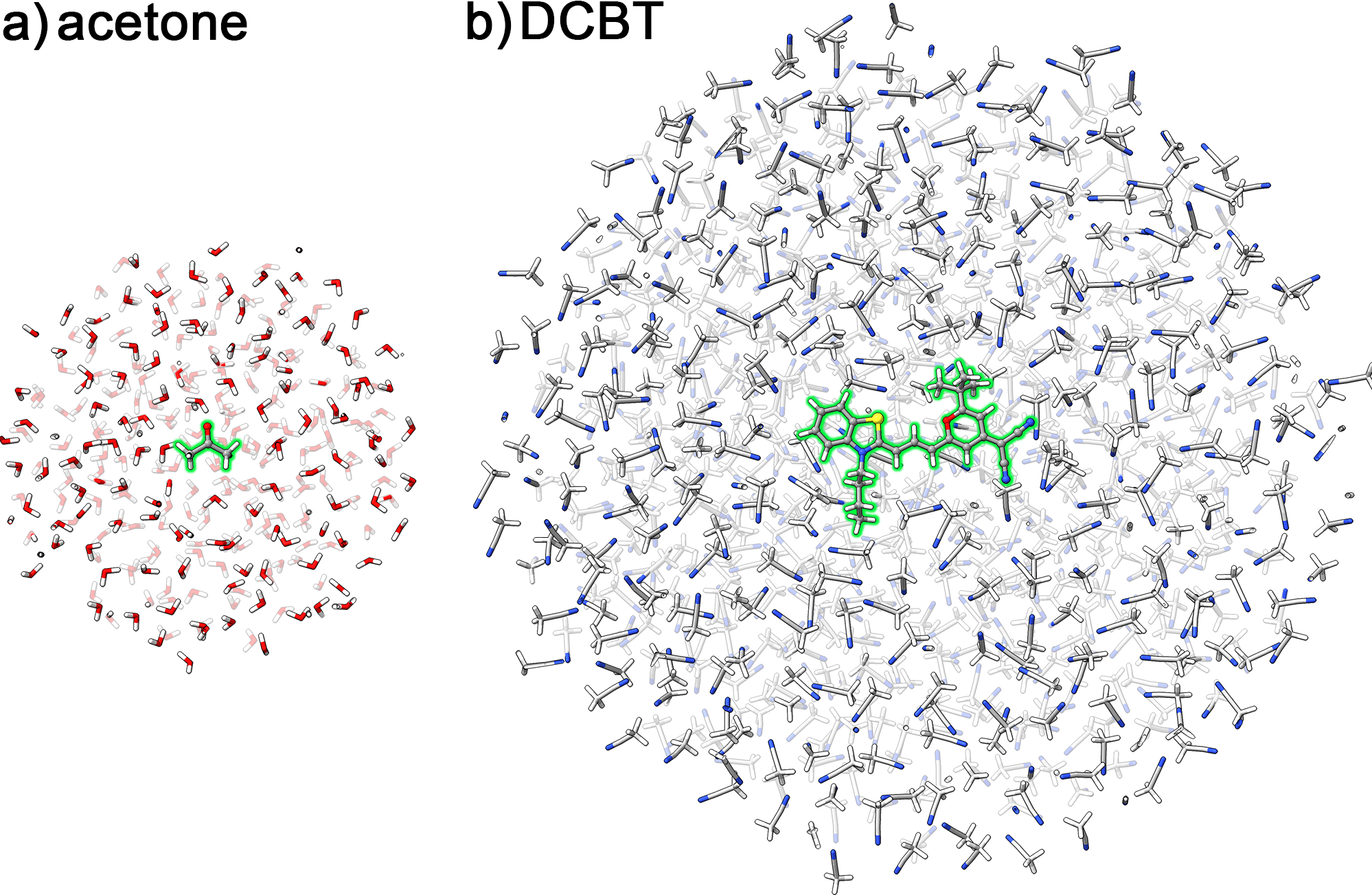} 
    \centering
    \caption{ Cross sections of representative snapshots of (a) acetone in water and (b) DCBT in acetonitrile. Solutes highlighted in green.}
    \label{f:plot_acetone_and_DCBT}
\end{figure}\textbf{}

\item \textit{QM/MM calculations}: Vertical excitation energies were computed for each configuration using the DMRG/FQ approach implemented in a locally modified version of OpenMolcas\cite{aquilante2020modern,li2023openmolcas}. 
For acetone, a full-valence (24,22) active space and aug-cc-pVDZ basis set were employed, and two different FQ parametrizations considered: FQ$^a$ from ref.~\citenum{rick1994dynamical}, and FQ$^b$ from ref.~\citenum{ambrosetti_parametrization}. To assess the influence of solute–solvent polarization, additional non-polarizable ESPF calculations~\cite{ferre2002approximate} employing TIP3P~\cite{jorgensen1981quantum} charges were performed. Moreover, CASSCF/FQ$^{a,b}$(12,10) calculations were carried out to evaluate the effect of expanding the active space from (12,10) to a full-valence one on the final results. For DCBT, only the DMRG/FQ$^b$ approach was employed with a (30,27) active space and the 6-31G* basis set.

Starting orbitals were generated at the HF/FQ level, followed by Pipek-Mezey localization.\cite{Pipek_Mezey} MOs were selected to define the active space with active orbitals arranged according to Fiedler vector ordering \cite{fiedler_1,fiedler_2,Legeza_2011}. SS-DMRG/FQ calculations were carried out for both the ground state (GS) and the first singlet excited state (ES). For computational efficiency, initial calculations were performed for both states with a maximum bond dimension of $M=100$. Orbitals obtained from the GS calculation were used as the initial guess of the ES calculation. Subsequently, the optimized orbitals of each state were used as starting orbitals for a more refined calculation, increasing the maximum bond dimension to $M=300$, the results of which were used to compute the vertical excitation energies. The Cholesky MEDIUM option in OpenMolcas was used for acetone and the RICD option for DCBT. 


\item \textit{Analysis and refinement}:
For each system, the excitation energy in solution was calculated by averaging the excitation energies over all snapshots. The solvatochromic shift was then obtained by subtracting the excitation energy in solution from the excitation energy in the gas-phase. The vertical excitation energies in the gas phase were computed for single structures optimised in the gas phase: 
the geometry of acetone was obtained from ref. \citenum{skoko}; the geometry of DCBT was optimized with Gaussian16\cite{g16} at the MP2/6-31G* level of theory.

\end{enumerate}

To validate the performance of the DMRG/FQ model, benchmarking was carried out on a single structure of acetone with two water molecules hydrogen-bonded to the carbonyl oxygen (see \cref{f:plot_acetone_2_water}). A series of basis sets (6-31G*, cc-pVDZ, and aug-cc-pVDZ),  active spaces [(4,3), (6,5), (12,10), and (24,22)], and solvation models (ESPF, FQ$^a$, and FQ$^b$) were examined. Note that for the smaller active spaces, sufficient values of $M$ were selected corresponding to the dimension of the active spaces: $M = 100$ for the (4,3) and (6,5) cases, and $M = 200$ for (12,10), while $M = 300$ for (24,22). For the active spaces up to (12,10), the HF orbitals were used as the initial guess for both the GS and the ES DMRG/FQ calculations whereas for (24,22) the protocol reported above in point 4 of the list was employed. FQ parameters for acetonitrile were taken from Ref.\citenum{ambrosetti_parametrization}

\section{Results and discussion}\label{sec:results}

\subsection{Model Validation}\label{sec:benchmark}

\begin{figure}[ht!]
 \includegraphics[scale=0.25]{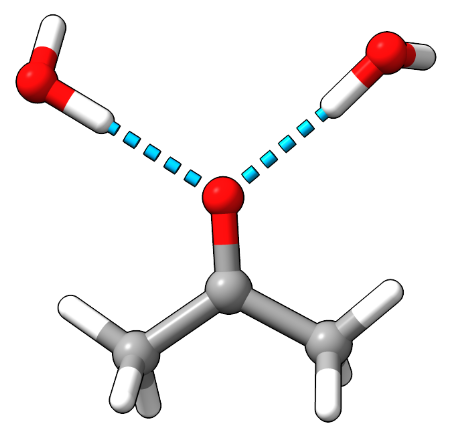} 
    \centering
    \caption{Representative structure of the model system that is exploited to validate DMRG/FQ. Acetone is treated at DMRG level, while two hydrogen-bonded water molecules are treated with FQ. Hydrogen bonds are illustrated with blue dashed lines.}
    \label{f:plot_acetone_2_water}
\end{figure}

To validate DMRG/FQ, the $n \rightarrow \pi^*$ excitation energy of acetone in aqueous solution is taken as a reference. We selected acetone in aqueous solution as a test system, as it represents a well-established model widely used as a benchmark in the study of photochemical and photophysical phenomena, including photoreactivity and photochromism-related processes.\cite{Ma2013AcetoneWater,Aloisio2000PhotochemistryAcetone,Norrish1934Acetone,skoko} Owing to its simple molecular structure and well-characterized excited-state behavior, acetone in water provides a reliable reference system for assessing the accuracy and robustness of theoretical and computational approaches aimed at describing solvent effects and light–matter interactions.
In particular, a representative structure is considered in which two water molecules donate hydrogen bonds to the carbonyl oxygen of acetone, as shown in \cref{f:plot_acetone_2_water}. For this model structure, the excitation energy is computed across a range of active spaces, basis sets, and solvent models. Gas-phase results, obtained with the same basis set, active space, and starting orbitals, will be taken as reference.

\begin{figure}[ht!]
 \includegraphics[width=\textwidth] {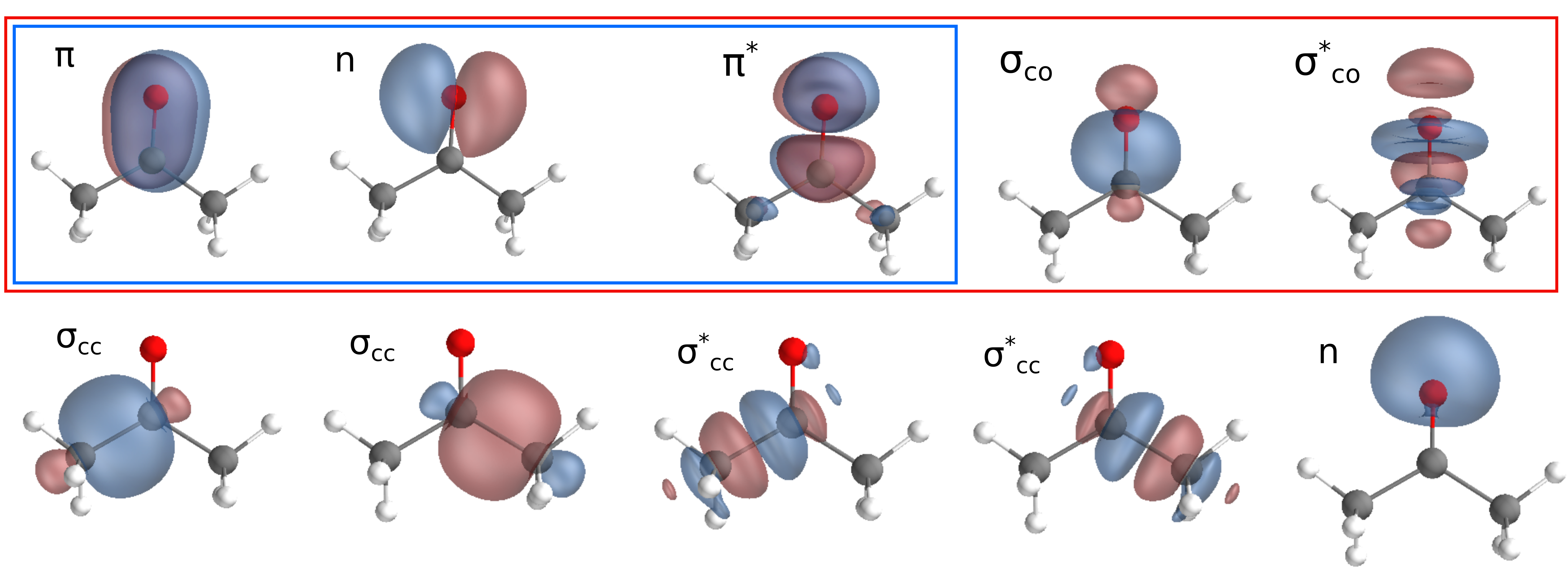} 
    \centering
    \caption{HF localized valence orbitals of acetone, which are employed as the initial guess in the DMRG calculations. Orbitals in the blue box define the (4,3) active space; those in the red box define the (6,5) space; the full set displayed corresponds to the (12,10) space. 
    }
    \label{fig:acetone_orbitals}
\end{figure}

The (4,3), (6,5), (12,10), and (24,22) active spaces are explored. The HF orbitals defining the first three active spaces are shown in \cref{fig:acetone_orbitals} while (24,22) corresponds to the full-valence space. 

\begin{figure}[hbt!]
  \includegraphics[scale=1.0]{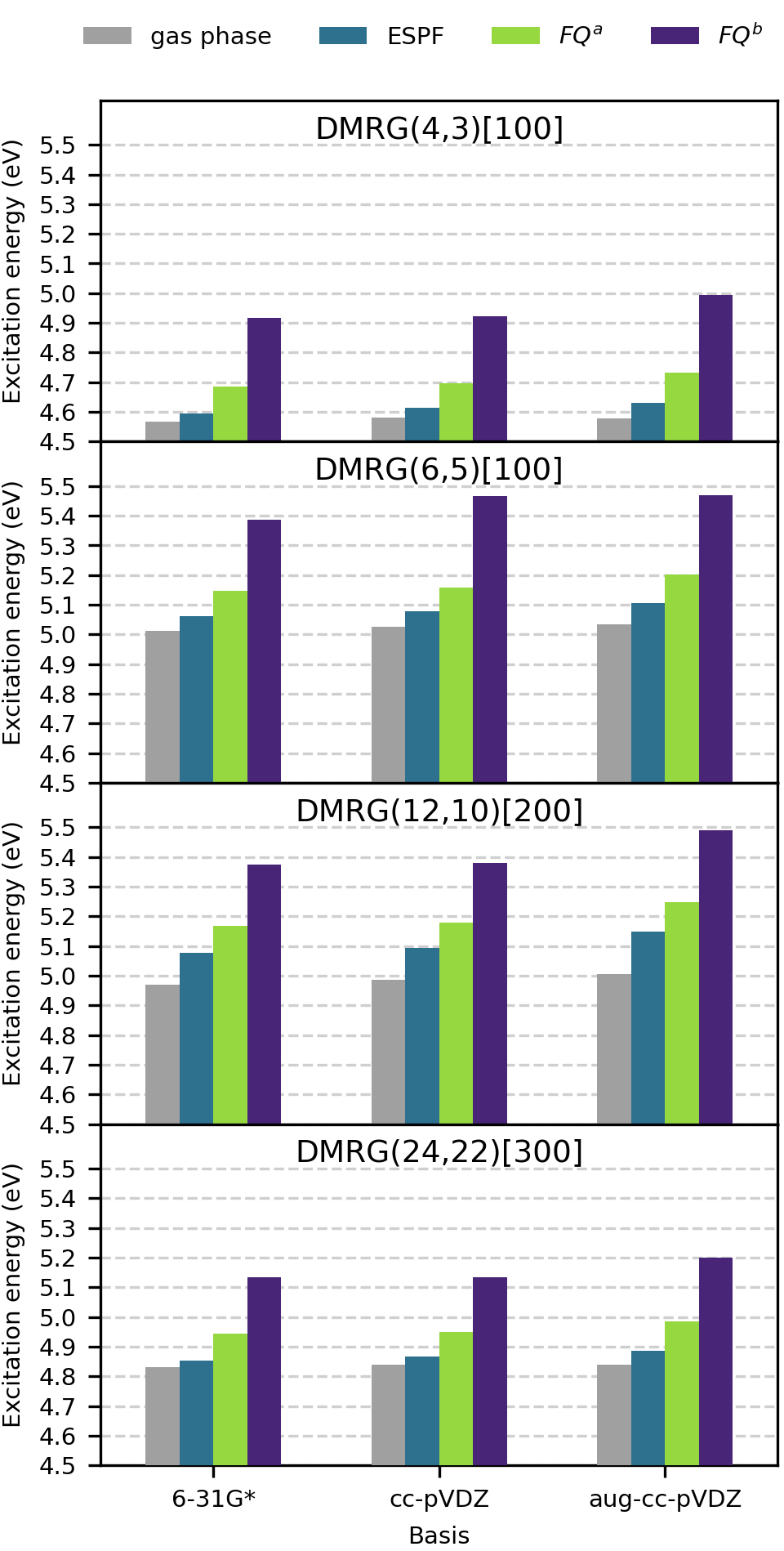} 
    \centering
    \caption{Computed $n \rightarrow \pi^*$ vertical excitation energies (eV) of acetone in the gas phase and hydrogen-bonded to two water molecules (single structure) for selected basis sets (6-31G*, cc-pVDZ, and aug-cc-pVDZ), active spaces [(4,3), (6,5), (12,10), and (24,22)], and solvation models (ESPF, FQ$^a$, and FQ$^b$). The number in square brackets represents $M$, the maximum bond dimension. }
    \label{fig:exc_en_vs_basis}
\end{figure}

The (4,3) active space includes four electrons in the $n$ orbital antisymmetric with respect to the plane perpendicular to the carbon skeleton, together with the $\pi$ and $\pi^*$ orbitals. The (6,5) active space is obtained by adding the carbonyl $\sigma$ and $\sigma^*$ orbitals to the (4,3) set. Further expansion leads to the (12,10) active space, which  includes the $\sigma$ and $\sigma^*$ orbitals of each C–C bond, as well as the symmetric $n$ orbital.
Finally, the (24,22) (full-valence) space includes the remaining six $\sigma$ and six $\sigma^*$ orbitals associated with the C–H bonds. 
The lowest values of the computed excitations were obtained with the (4,3) active space (see \cref{fig:exc_en_vs_basis} and table S1 in the SI). As the active space expands to (6,5) and (12,10) excitation energies generally increase; however, a decrease can be observed for the full-valence (24,22) space. This behavior reflects the increasing stabilization of the ES relative to the GS as the active space approaches the full-valence limit.

To assess the impact of polarization and diffuse functions on the computed excitation energies, the 6-31G*, cc-pVDZ, and aug-cc-pVDZ basis sets were employed. In addition, the solvent environment was described using the non-polarizable ESPF approach \cite{ferre2002approximate} and the polarizable FQ model for two different parametrizations (FQ$^a$ from ref.~\citenum{rick1994dynamical} and FQ$^b$ from ref.~\citenum{ambrosetti_parametrization}). Regarding the basis set, a systematic increase in the calculated excitation energies can be seen, moving from 6-31G* to cc-pVDZ and further to aug-cc-pVDZ (see \cref{fig:exc_en_vs_basis} and table S1 in the SI). The effect is most pronounced for FQ$^b$, the solvent parametrization that yields the highest excitation energies.
All excitation energies calculated in solution are larger than the corresponding gas-phase values, indicating a solvent-induced blue shift. This solvatochromic shift increases when moving from the ESPF model, which uses fixed TIP3P charges, to the FQ$^a$, and subsequently FQ$^b$ models. This increase can be explained by considering the different parametrizations of the solvent approaches: ESPF with TIP3P charges \cite{ferre2002approximate} and FQ$^a$ \cite{rick1994dynamical} are designed to reproduce bulk water properties, with FQ$^a$ additionally accounting for solute-solvent polarization. In contrast, FQ$^b$ targets solute-solvent electrostatic and polarization interactions\cite{ambrosetti_parametrization} leading to a stronger solvent effect.
Evidently, the choice of the active space and of the solvent model (and its parametrization) plays a major role in determining the computed excitation energies. 

To end this discussion, it is important to note that for the (4,3) and (6,5) active spaces, there is an inconsistency between GS and ES optimized active orbitals. In the ES, the orbital with $n$ character is antisymmetric with respect to the symmetry plane perpendicular to the carbon skeleton, as expected. In the GS, however, the corresponding orbital is symmetric with respect to the same plane (see fig. S2 in the SI). To address this problem and enforce consistency between GS and ES orbitals, GS calculations using (2,2) and (4,4) active spaces were performed and compared to ES calculations at the (4,3) and (6,5) levels, respectively (see table S1 and fig. S3 in the SI). In this way, the symmetric $n$ orbital is always kept in the core of the DMRG calculation.  
All excitation energies for these active spaces are reduced by about 0.05 eV compared to those calculated with the (4,3) and (6,5) active spaces for both the GS and the ES.  

\subsection{Acetone in aqueous solution}

\begin{figure}[ht!]
 \includegraphics[width=0.4\textwidth]{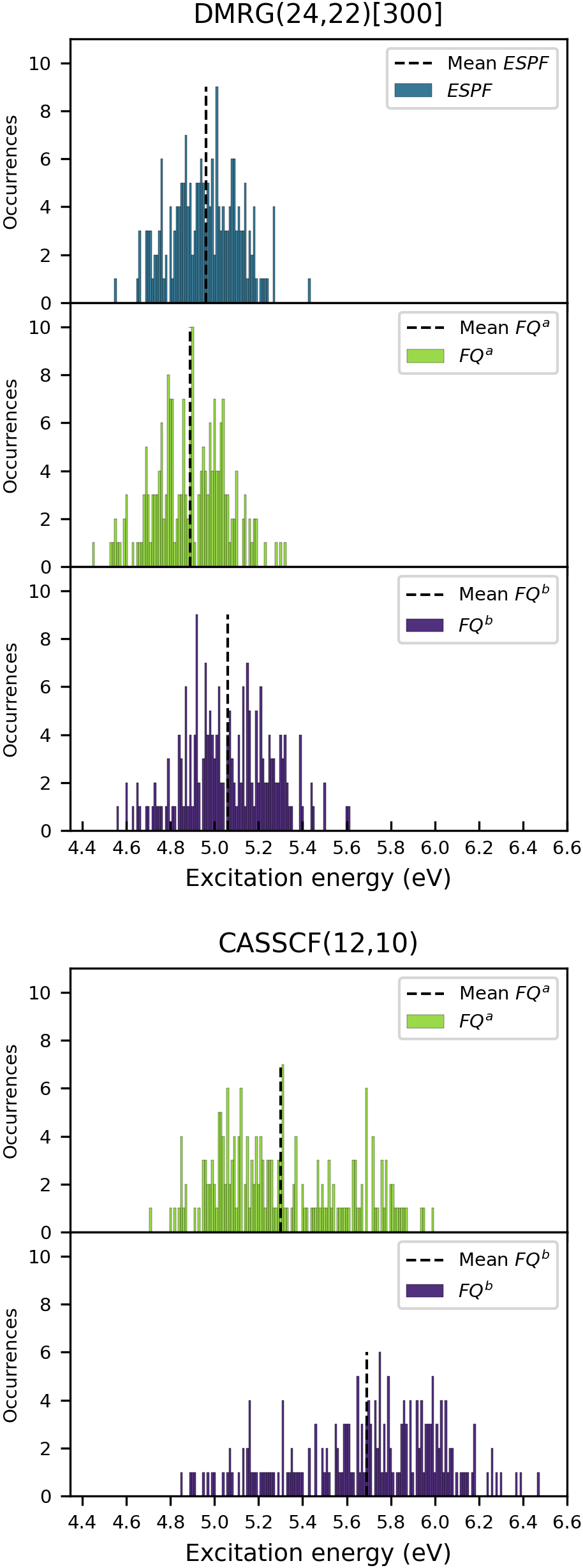} 
    \centering
    \caption{Distributions of vertical excitation energies (eV) computed for the $n \rightarrow \pi^*$ transition of acetone in aqueous solution using (top) DMRG/ESPF(24,22), DMRG/FQ$^a$(24,22), and DMRG/FQ$^b$(24,22) compared to (bottom)  CASSCF/FQ$^a$(12,10) and CASSCF/FQ$^b$(12,10). All values refer to 200 snapshots. The mean excitation energies are indicated by black dashed lines. [300] refers to the maximum bond dimension $M$.}
    \label{fig:hist_exc_en_dmrg}
\end{figure}

Based on the validation reported above, in this section DMRG/FQ is applied to simulate the $n \rightarrow \pi^*$ excitation of acetone in aqueous solution, according to the protocol reported in Section\ref{sec:comp_det}. Absorption energies are computed for 200 snapshots extracted from the MD trajectory using the polarizable DMRG/FQ$^{a,b}$ levels of theory and compared to non-polarizable DMRG/ESPF and CASSCF(12,10)/FQ$^{a,b}$ calculations (see \cref{fig:hist_exc_en_dmrg}). The comparison with DMRG/ESPF 
assesses the effect of mutual solute-solvent polarization, while the comparison with CASSCF(12,10)/FQ aims to evaluate the impact of expanding the active space from (12,10) to the full valence limit. All calculations are performed using the aug-cc-pVDZ basis set and localized HF starting orbitals, with the (24,22) full valence active space and $M=300$ for DMRG in accordance with the validation described in \cref{sec:benchmark}. 
Convergence with respect to the number of frames is evaluated by calculating the average excitation energies over the first 50, 100, and 150 snapshots out of the total 200, confirming that 200 snapshots are sufficient for reliable convergence (table S4).


The calculated absorption energies show large fluctuations across the snapshots due to variations in the solute conformations (see \cref{fig:hist_exc_en_dmrg}) and the dynamic behavior of the water molecules surrounding acetone.
This broadening is dependent on the solvent model used. Specifically, the spread of excitation energy is 0.88 eV 
for DMRG/ESPF, 0.87 eV for DMRG/FQ$^a$, and 1.05 eV for DMRG/FQ$^b$ while the mean excitation energies are 4.96 eV, 4.89 eV, and 5.06 eV, respectively (see \cref{fig:hist_exc_en_dmrg} and table S6). 
For a full report of mean, median, mode, and standard error, refer to table S2 in the SI. These results highlight how different atomistic approaches provide distinct descriptions of solute-solvent interactions; however, the difference between mean and median for all solvent models is approximately 0.01 eV, suggesting nearly symmetric distributions in all cases. 

The results obtained with the CASSCF/FQ(12,10) calculations (see \cref{fig:hist_exc_en_dmrg} and table S6) show a larger spread of energy: 1.28 eV for CASSCF/FQ$^a$ (mean = 5.30 eV), and 1.62 eV for CASSCF/FQ$^b$ (mean = 5.69 eV). This broader distribution likely reflects the variability in the active orbitals, which results from their incompleteness relative to the full valence case. Table S2 in the \sm also presents the mean, median, mode, and standard error of the mean for the excitation energies for the CASSCF/FQ(12,10) level of theory. 

\begin{figure}[ht!]
 \includegraphics[scale=1.0]{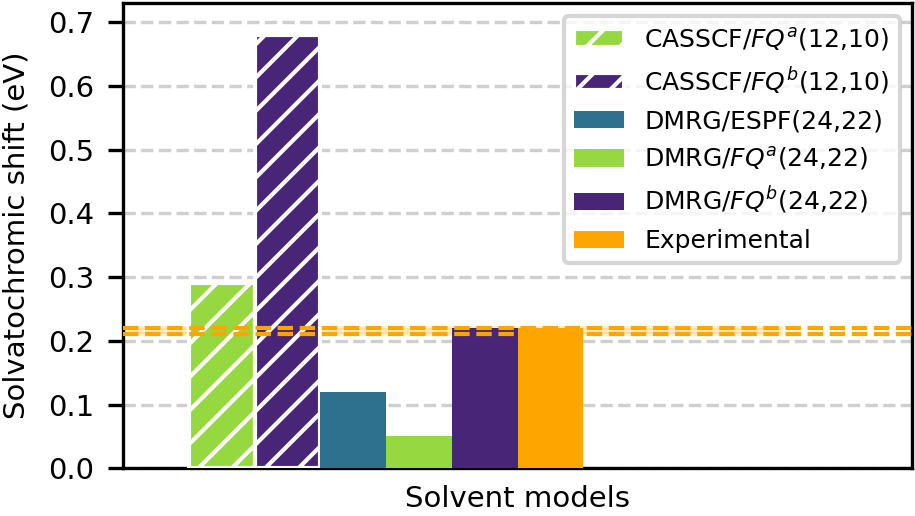} 
    \centering
    \caption{Computed water-to-vacuum solvatochromic shifts of the $n \rightarrow \pi^*$ excitation of aqueous acetone using DMRG/ESPF(24,22), DMRG/FQ$^a$(24,22), and DMRG/FQ$^b$(24,22) compared to CASSCF/FQ$^a$(12,10) and CASSCF/FQ$^b$(12,10) methods, together with the experimental value \cite{renge2009solvent,bayliss1968solvent}.}
    \label{fig:plot_solv_shift_acetone}
\end{figure}

For water-to-vacuo solvatochromic shifts, all models yield a blue shift, the largest for the FQ$^b$ solvent model (see \cref{fig:plot_solv_shift_acetone} and table S6). 
Compared to experimental values (0.22 eV\cite{renge2009solvent} and 0.21 eV\cite{bayliss1968solvent}), DMRG/ESPF(24,22) and DMRG/FQ$^a$(24,22) underestimate the solvatochromic shift, giving 0.12 eV and 0.05 eV, respectively. DMRG/FQ$^b$(24,22), however, yields 0.22 eV, consistent with experiment. In contrast, CASSCF/FQ$^{(a,b)}$(12,10) overestimates the shift (0.29 and 0.68 eV), highlighting the advantage of a full-valence active space over the smaller (12,10). These findings demonstrate that combining a full valence active space calculation —prohibitively large for conventional CASSCF—with the FQ$^b$ parametrization (which, as already reported above, is tailored to reproduce solute-solvent polarization) provides the most reliable description of the solvated system among those tested. 


It is worth noticing that, while the computed excitation energies qualitatively reproduce the experimental trends, the absolute values in both the gas phase and aqueous solution remain larger than experimental values due to the lack of dynamic electron correlation in the DMRG and CASSCF calculations (table S6). 
The effect of dynamic correlation can be estimated using the CASPT2 approach \cite{andersson1990second} to improve quantitative accuracy. Gas-phase CASPT2 calculations with the (12,10) active space yield an excitation energy of 4.46 eV, in good agreement with the experimental data reported in table S6. The dynamic correlation contribution, relative to CASSCF(12,10) (which yields an excitation energy of 5.01 eV), amounts to 0.55 eV.

Additionally, the DMRG/FQ coupling completely neglects solute–solvent non-electrostatic interactions. In particular, we have recently shown that solute-solvent Pauli repulsion is particularly relevant for consistently modeling vacuo-to-water solvatochromic shifts \cite{giovannini2019quantum}. Pauli repulsion is expected to confine the QM density, thereby reducing the absolute value of the solvatochromic shift.\cite{egidi2021polarizable,giovannini2019quantum,Amovilli_Floris_rep_2019,Amovilli_Floris_rep_2020} 
The inclusion of dynamic correlation and non-electrostatic terms within the QM/FQ framework is therefore expected to provide quantitatively accurate excitation energies and refine the computed solvatochromic shifts.

\subsection{DCBT in acetonitrile}


To demonstrate the applicability of the method to larger systems, DMRG/FQ is applied to the simulation of the bright $\pi \rightarrow \pi^*$ excitation of DCBT \cite{paper_DCBT} in acetonitrile. 

\begin{figure}[ht]
\includegraphics[scale=1.0]{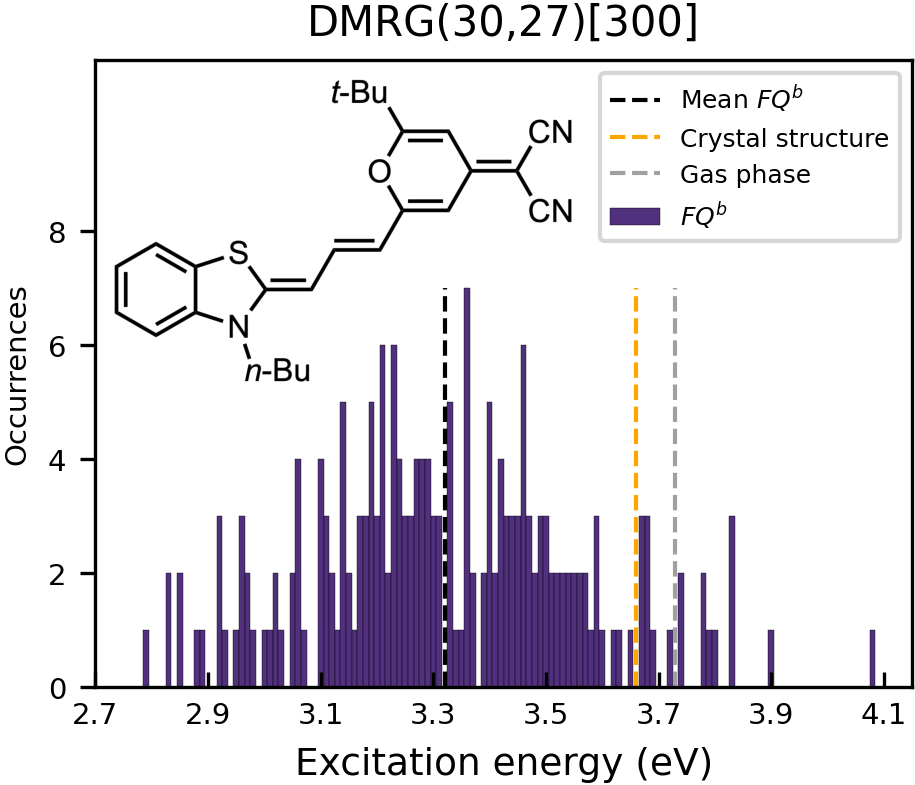} 
    \centering
    \caption{Distribution of computed $\pi \rightarrow \pi^*$ transition energies (eV) of DCBT in acetonitrile using DMRG/FQ$^b$. The purple dashed line marks the mean excitation energy. The grey and orange dashed lines indicate the gas-phase values calculated using the optimized structure and crystal structure reported in Ref. \citenum{paper_DCBT}, respectively.}
    \label{fig:hist_exc_en_DCBT}
\end{figure}

DCBT is a push–pull merocyanine dye featuring a strongly conjugated donor–acceptor architecture (see Fig.\ref{fig:hist_exc_en_DCBT}). Its main applicative interest lies in its environment-dependent fluorescence behavior, particularly the pronounced sensitivity of its emission efficiency to solvent polarity and local molecular surroundings.\cite{paper_DCBT} DCBT serves as a valuable model system for investigating non-radiative decay pathways and excited-state dynamics, enabling a deeper understanding of how molecular structure and environment govern fluorescence quantum yields. These insights are crucial for the rational design of high-performance fluorophores with controlled emission properties.

In line with the previous section, DCBT absorption energies are computed on 200 snapshots at the DMRG/FQ$^b$ level and the convergence assessed as described above for acetone (see table S5 in the SI). 
All calculations are performed using the 6-31G* basis set and localized HF starting orbitals. The (30,27) active space is employed, which includes all $\pi$ orbitals orthogonal to the molecular plane, as well as the $\pi$ orbitals of the C–N bonds lying in the plane. For each $\pi$ orbital, the corresponding $\pi^*$ orbital is also included, except for those associated with lone pairs localized on heteroatoms O, S and N. Excitation energies are obtained using a maximum bond dimension $M=300$. 

\begin{figure}[ht!]
 \includegraphics[scale=1.0]{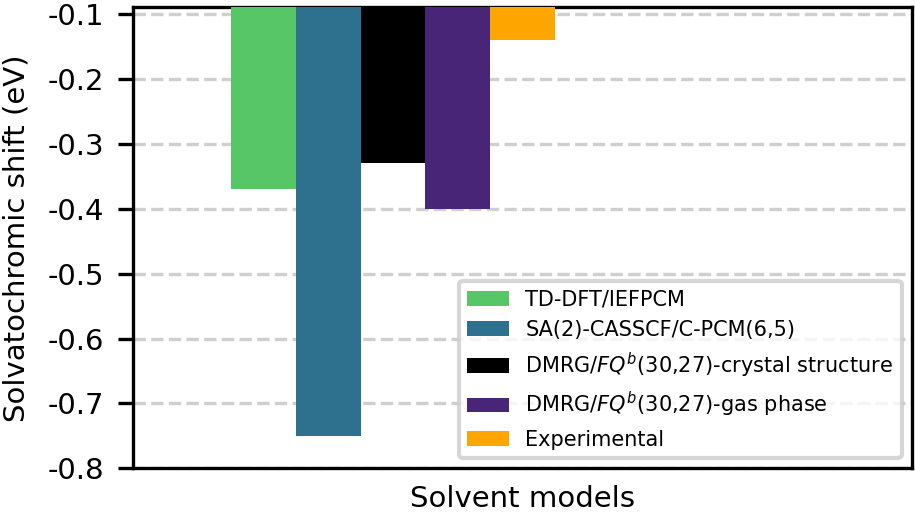} 
    \centering
    \caption{Computed acetonitrile-to-vacuum solvatochromic shifts of the $\pi \rightarrow \pi^*$ excitation of DCBT .  DMRG/FQ$^b$(30,27) values obtained by computing the gas phase reference at two different structures (see text) are shown, together with TD-DFT/IEFPCM and SA(2)-CASSCF/C-PCM(6,5)\cite{song_casscf_sa_pcm} values. The experimental and TD-DFT/IEFPCM values are obtained by taking methylcyclohexane as a proxy for the gas-phase (see text)\cite{paper_DCBT}.}
    \label{fig:plot_solv_shift_dcbt}
\end{figure}

As with acetone in aqueous solution, the computed DMRG/FQ$^b$ absorption energy for DCBT in acetonitrile varies substantially from snapshot to snapshot, reflecting the role of the different geometrical arrangements of water around DCBT (see \cref{fig:hist_exc_en_DCBT}).
Specifically, the excitation energies span 1.29 eV around a mean of 3.32 eV (from 2.79 eV to 4.08 eV). 
The values of the mean, median, mode, and standard error of the mean of the excitation energies are reported in table S3 of the \sm. 

To compute the solvent-to-vacuum solvatochromic shift, gas-phase references are taken from the excitation energies of (i) the optimized structure (as reported in \cref{sec:comp_det}) and (ii) the crystal structure reported in ref. \citenum{paper_DCBT}, namely 3.73 eV and 3.66 eV, respectively. These references yield shifts of -0.41 and -0.34 eV, both indicating a red shift.

These results can be compared with computed TDDFT/IEFPCM values and experimental data reported in Ref.\citenum{paper_DCBT}, as well as with state-averaged SA(2)-CASSCF/C-PCM(6,5) values \cite{song_casscf_sa_pcm} (see Figure S4 in the \sm\ for more details on how these data were extracted).

Experimental gas-phase spectra of DCBT are not available in the literature. Therefore, experimental spectra in methylcyclohexane can be exploited as a proxy for the gas-phase, because this solvent is the one with the lowest dielectric constant ($\epsilon_r = 2.02$) among those measured in Ref. \citenum{paper_DCBT}. Under these conditions, the experimental spectrum yields a shift of -0.15 eV, while TDDFT/IEFPCM (def2-TZVP) gives -0.38 eV (see also table S8 of the \sm for more details). Both values are in fair agreement with DMRG/FQ$^b$ values (-0.41/-0.34). In fact,  they indicate a red shift, which is correctly reproduced by DMRG/FQ$^b$, and are expected to underestimate the actual solvatochromic shift, as they correspond to DCBT in methylcyclohexane rather than in the gas phase.

In ref. \citenum{song_casscf_sa_pcm}, state-averaged SA(2)-CASSCF/C-PCM (6,5) calculations using the 6-31G* basis set were performed, yielding excitation energies of 4.44 eV in the gas phase and 3.68 eV in dimethyl sulfoxide. The latter value can be taken as a proxy for acetonitrile, given the similar dielectric constants of the two solvents (35.1 for acetonitrile and 46.7 for dimethyl sulfoxide) with PCM surface charges scaling as $\frac{\epsilon_r - 1}{\epsilon_r}$. The corresponding solvatochromic shift is –0.76 eV. This value appears to be overestimated, perhaps reflecting an excessively fast solvent response in PCM calculations. Indeed, according to the data of Ref.\citenum{song_casscf_sa_pcm}, the solvatochromic shift for toluene ($\epsilon_r = 2.38$) is already -0.40 eV (see also table S9 of the \sm for more details). Hence, our calculated solvatochromic shift, which lies between the values extracted from Ref.\citenum{paper_DCBT} and Ref.\citenum{song_casscf_sa_pcm} discussed above, confirms the reliability of our approach. Note that other potential sources of inaccuracy in PCM values include the smaller (6,5) active space that was employed and the shape and size of the molecular cavity.

Finally, the effect of dynamic correlation for DCBT in the gas phase can be estimated with CASPT2 \cite{andersson1990second}. A CASSCF(8,8) calculation—using an active space comprising the $\pi$ orbitals along the polymethine chain—yields an excitation energy of 3.94 eV. The subsequent CASPT2 calculation within the same active space gives 2.99 eV, indicating a dynamic-correlation lowering of -0.95 eV. Applying this estimated correction to the DMRG/FQ result leads to an excitation energy of about 2.37 eV, close to the experimental value of 2.27 eV. Note that this value is only an estimate of the dynamic correlation and could change slightly if the perturbative correction were applied to the DMRG/FQ$^b$(30,27) level of theory. As already mentioned above for acetone, also in this case, incorporating dynamic correlation explicitly in the QM/FQ framework—together with non-electrostatic interaction terms—should further improve the absolute excitation energies and the predicted solvatochromic shifts.

\section{Summary and conclusions}

In this work we presented an integrated DMRG/FQ framework for the simulation of solvated molecular systems, combining the accuracy of the Density Matrix Renormalization Group (DMRG) with the flexibility of the fluctuating‐charge (FQ) force field. The method exploits the MPS–MPO formulation of DMRG and its orbital‐optimization capabilities to capture static electron correlation in the quantum region, while the FQ model provides a polarizable classical environment.
The approach was validated on representative solute–solvent systems, including acetone in water and the DCBT chromophore in acetonitrile. Using extensive MD sampling, we demonstrated that DMRG/FQ reliably describes solvent-induced polarization and yields excitation energies and solvatochromic shifts that are in good agreement with experiment, particularly when the FQ$^b$ parametrization is employed. The observed spectral spreading and average excitation energies underline the method’s capability to capture the interplay between electronic structure and solvent fluctuations.

Overall, the DMRG/FQ scheme constitutes a significant step forward in the multiscale modeling of electronically excited states in complex environments. Future developments should focus on extending the framework to incorporate dynamic electron correlation, for instance through perturbative schemes\cite{Yanai_2011,QCMaquis_v_4,dmrg_nevpt2} or using DMRG-DFT hybrid approches \cite{DMRG-PDFT,dmrg_in_dft_veis}, and non-electrostatic solute–solvent interactions, which are expected to further improve absolute excitation energies and solvatochromic predictions \cite{giovannini2019disrep,amovilli_floris_disp_2023,amovilli_floris_int_2025}. Such enhancements will broaden the applicability of DMRG-based embedding methods to increasingly complex chemical and photochemical processes.

\begin{acknowledgement}
The authors acknowledge funding from MUR-FARE Ricerca in Italia: Framework per l'attrazione ed il rafforzamento delle eccellenze per la Ricerca in Italia - III edizione. Prot. R20YTA2BKZ and the European Union's Horizon Europe research and innovation programme under the project HORIZON-MSCA-2023-DN-01 - LUMIÈRE G.A . No 101169312. The Center for High-Performance Computing (CHPC) at SNS is also acknowledged for providing the computational infrastructure.
\end{acknowledgement}
\begin{suppinfo}
Additional computational details of molecular dynamics simulations, benchmarking and statistical summaries of calculated vertical excitiation energies, convergence of vertical excitations regarding number of snapshots, and digitized DCBT absorption spectra (PDF). 

\end{suppinfo}

\bibliography{biblio}

\end{document}


\tableofcontents

\newpage

\section{Molecular dynamics of DCBT in acetonitrile}\label{sec:md-dcbt}
\begin{figure}
    \centering
    \includegraphics[]{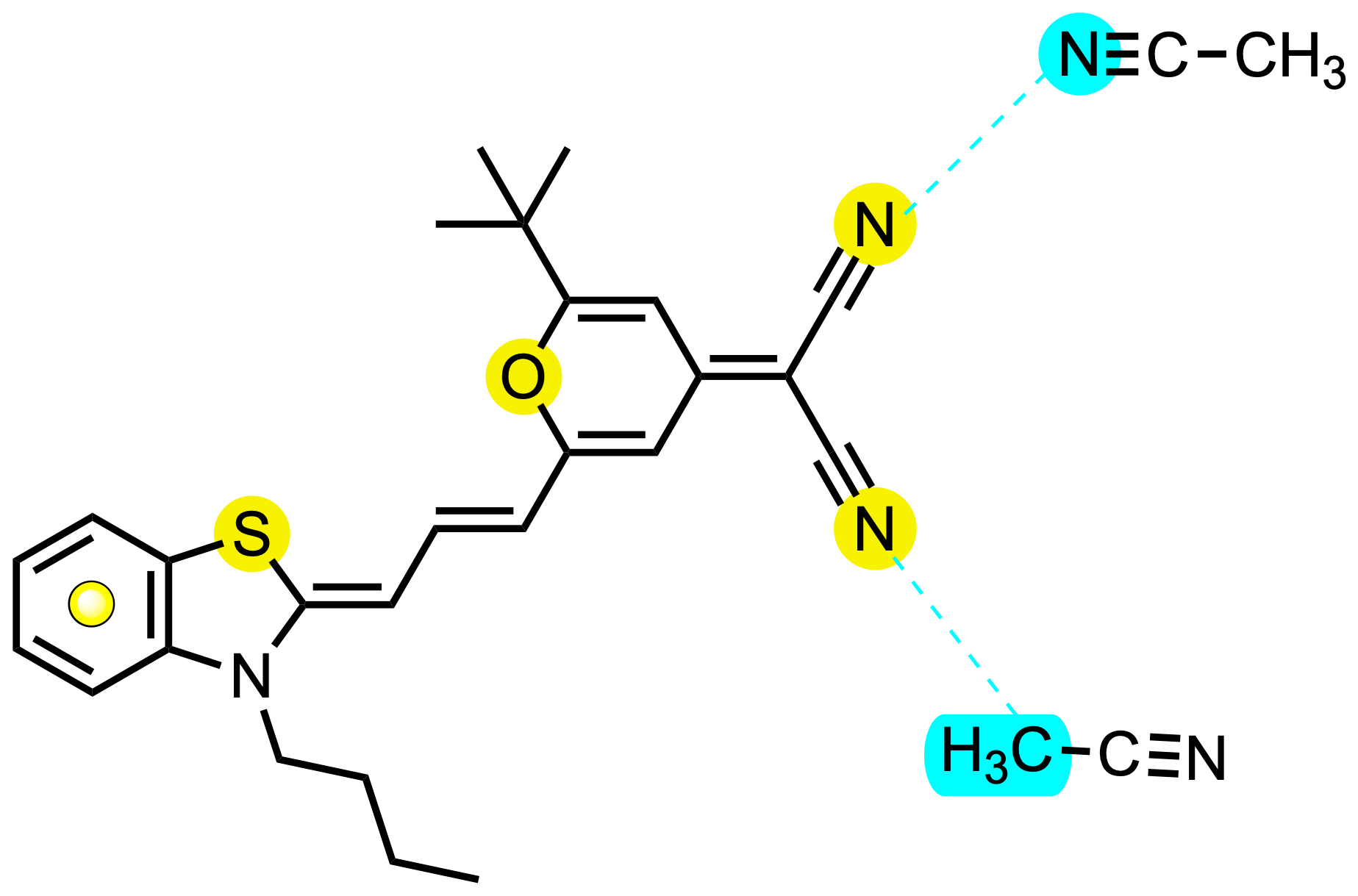}
    \includegraphics[width=0.8\linewidth]{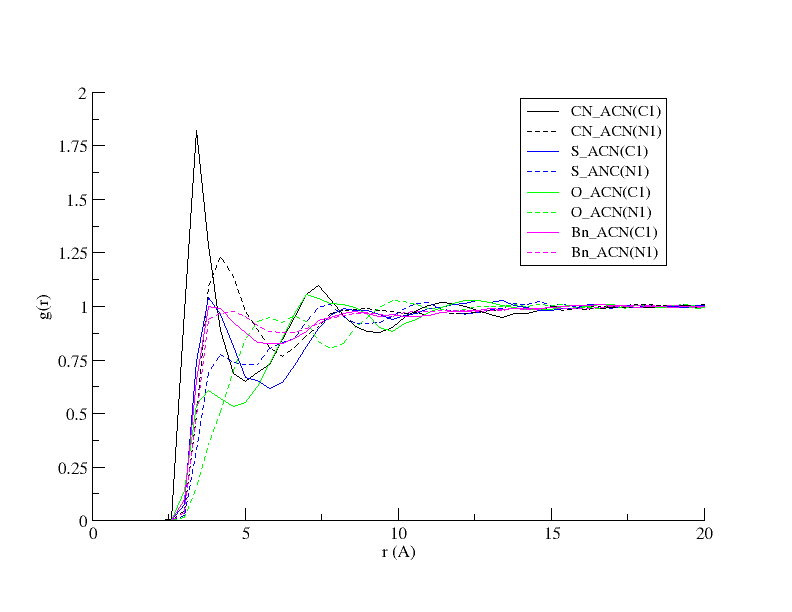}
    \caption{Radial distribution function (RDF) for selected DCBT and acetonitrile (ACN) atom pairs. Top diagram highlights in yellow DCBT atoms from which the RDF is measured; note Bn refers to the centre point of the benzene moiety; highlighted in blue are ACN(C1), which refers to the acetonitrile methyl carbon and ACN(N1), which refers to the acetonitrile nitrogen. Ordering of ACN methyl groups toward DBCT nitrile substituents (black lines) is evident.}
    \label{fig:RDFgraph}
\end{figure}

Molecular dynamics simulations were performed using GROMACS 2020.4\cite{abraham2015gromacs}. Parameters for DCBT were generated with ACPYPE-\textit{antechamber}\cite{acpype,antechamber} using the General Amber Force Field (GAFF),\cite{gaff} AM1-BCC charges,\cite{am1bcc} and molecular geometry optimised prior using Gaussian16 at the MP2/6-31G(d) level of theory with implicit acetonitrile incorporated using the Polarizable Continuum Model (PCM).\cite{g16,mp2a,mp2b,pcm} Parameters for a six-point gaff-derived acetonitrile model were sourced from the work of Kowsari and co-workers \cite{kowsari}. DCBT was solvated with approximately 17000 acetonitrile molecules in a cubic box of 11.54 nm. The system was minimised using steepest descent. Short-range electrostatic and Van de Waals cut-offs were set to 1.2 nm and long range electrostatic interactions were treated using particle-mesh Ewald (PME) with periodic boundary conditions.\cite{pme} Throughout all stages, strong position restraints (10000 kJ/mol/nm²) were applied to DCBT to maintain the planarity of the conjugated core while the butyl side chains were free to move. Using an NVT ensemble, the system was heated and equilibrated to 298.15 K for 1 ns using the velocity-rescaling method (0.1 ps coupling constant) and time step of 2 fs. All bonds were constrained using the LINCS algorithm.\cite{LINCSa,LINCSb} The system density was then equilibrated for 2ns under the NPT ensemble using the Berendsen barostat (2 ps coupling constant). Returning to the NVT ensemble, a final production run was performed for 30 ns from which 200 uncorrelated snapshots were extracted at regular 150 ps intervals. The MDanalysis program was used to calculate the radical distribution function of the solvent (Figure \ref{fig:RDFgraph}).\cite{mdanalysis}

\newpage

\section{Benchmarking of DMRG/FQ excitation energies}\label{sec:sec_exc_en}

\begin{table}[h!] 
    \caption{Computed $n \rightarrow \pi^*$ vertical excitation energies of acetone with two water molecules hydrogen-bonded to the carbonyl oxygen (single structure), for a range of basis sets, active spaces, and solvation models. For the (4,3) and (6,5) active spaces, “GS (2,2)” denotes the use of the (2,2) active space for the GS and the (4,3) active space for the ES, while “GS (4,4)” denotes the (4,4) active space for the GS and the (6,5) active space for the ES.}
    \label{t:exc_en_valid}
    \centering
    \setlength{\tabcolsep}{4pt}
    \renewcommand{\arraystretch}{1.2}
    \begin{tabular}{|c|c|c|cc|cc|cc|cc|}
    \hline
    \multirow{3}{*}{Active space} & \multirow{3}{*}{M} & \multirow{3}{*}{Basis set} 
        & \multicolumn{8}{c|}{Solvent model} \\
    \cline{4-11}
     &  &  & \multicolumn{2}{c|}{Gas-phase} 
        & \multicolumn{2}{c|}{ESPF} 
        & \multicolumn{2}{c|}{FQ$^a$} 
        & \multicolumn{2}{c|}{FQ$^b$} \\
    \cline{4-11}
     &  &  & & GS(2,2) & & GS(2,2) & & GS(2,2) & & GS(2,2) \\
    \hline
    (4,3) & 100 & 6-31g*       & 4.57 & 4.52 & 4.60 & 4.55 & 4.69 & 4.64 & 4.92 & 4.87 \\
     &  & cc-pVDZ      & 4.58 & 4.53 & 4.61 & 4.57 & 4.70 & 4.65 & 4.92 & 4.88 \\
     &  & aug-cc-pVDZ  & 4.58 & 4.53 & 4.63 & 4.59 & 4.73 & 4.69 & 4.99 & 4.95 \\
    \hline
    \cline{4-11}
     &  &  & & GS(4,4) & & GS(4,4) & & GS(4,4) & & GS(4,4) \\
    \hline
    (6,5) & 100 & 6-31g*       & 5.01 & 4.94 & 5.06 & 5.00 & 5.15 & 5.08  & 5.39 & 5.32 \\
     &  & cc-pVDZ      & 5.03 & 4.96 & 5.08 & 5.02 & 5.16 & 5.09 & 5.47 & 5.40 \\
     &  & aug-cc-pVDZ  & 5.03 & 4.97 & 5.11 & 5.04 & 5.20 & 5.14 & 5.47 & 5.41 \\
    \hline
    (12,10) & 200 & 6-31g*     & 4.97 & --   & 5.08 & --   & 5.17 & --   & 5.37 & --   \\
     &  & cc-pVDZ    & 4.99 & --   & 5.10 & --   & 5.18 & --   & 5.38 & --   \\
     &  & aug-cc-pVDZ& 5.01 & --   & 5.15 & --   & 5.25 & --   & 5.49 & --   \\
    \hline
    (24,22) & 300 & 6-31g*     & 4.83 & --   & 4.85 & --   & 4.94 & --   & 5.13 & --   \\
    &  & cc-pVDZ    & 4.84 & --   & 4.87 & --   & 4.95 & --   & 5.13 & --   \\
     &  & aug-cc-pVDZ& 4.84 & --   & 4.89 & --   & 4.99 & --   & 5.20 & --   \\
    \hline
    \end{tabular}
\end{table}

\newpage

\begin{figure}[ht!]
 \includegraphics[scale=0.5]{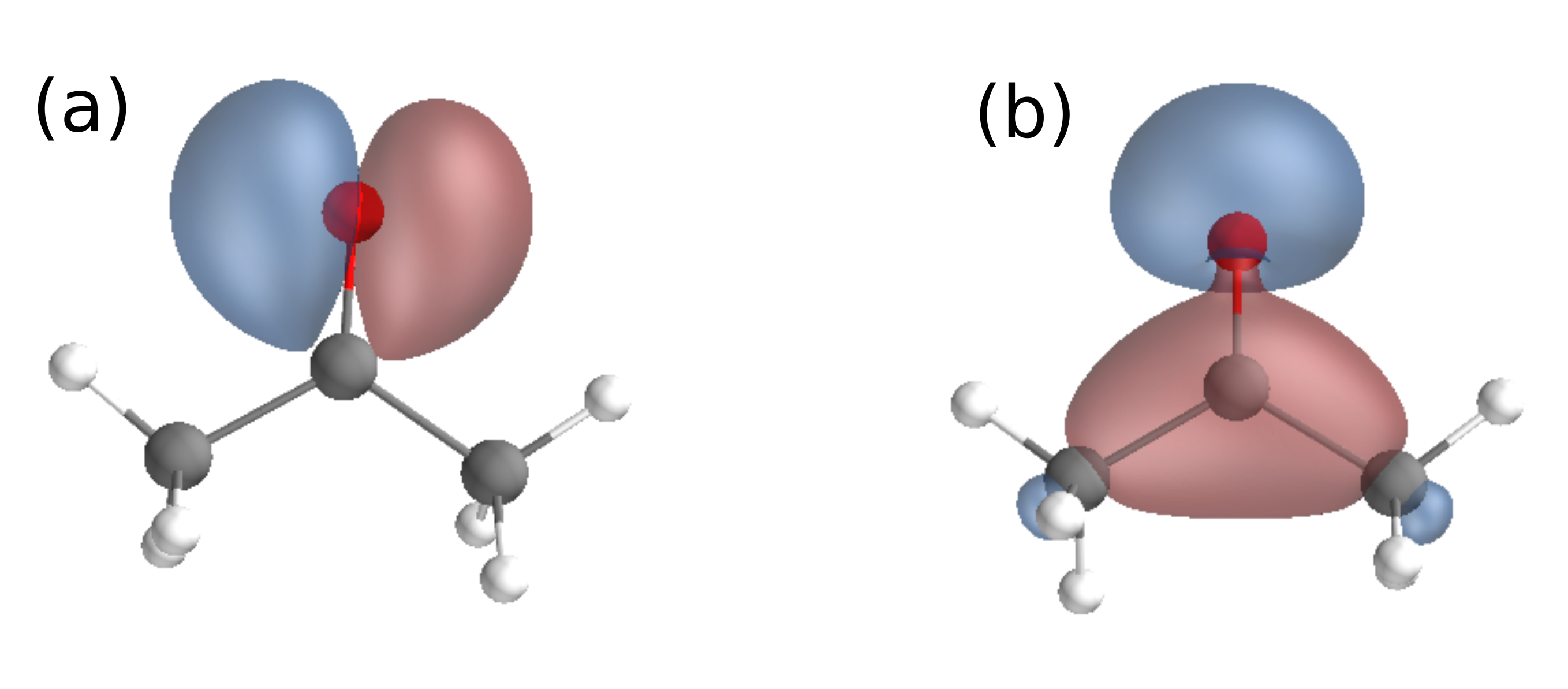} 
    \centering
    \caption{ \textbf{a)} 'Antisymmetric' $n$ orbital included in the (4,3) and (6,5) active spaces. \textbf{b)} 'Symmetric' orbital of  $n$ character, replacing the 'antisymmetric' one in the (4,3) and (6,5) active spaces after GS optimization.}
    \label{fig:n-s_and_n-a}
\end{figure}

\newpage

\begin{figure}[ht!]
 \includegraphics[width=\textwidth]{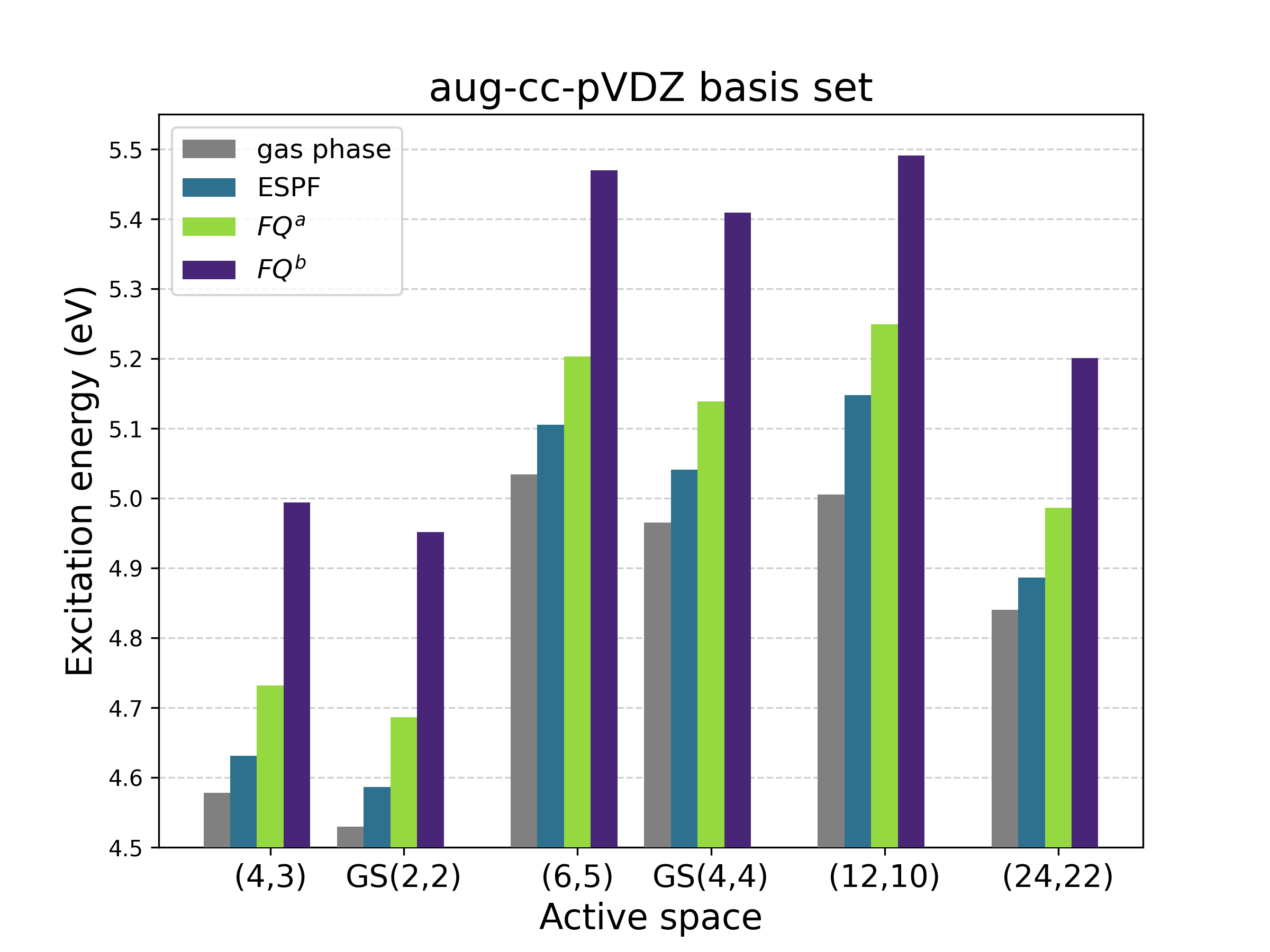} 
    \centering
    \caption{Computed $n \rightarrow \pi^*$ vertical excitation energies (eV) of acetone hydrogen bonded to two water molecules (single structure) using the aug-cc-pVDZ basis set. Results are compared for selected, active spaces and solvation models. “GS (2,2)” denotes the use of the (2,2) active space for the GS and the (4,3) active space for the ES, while “GS (4,4)” denotes the (4,4) active space for the GS and the (6,5) active space for the ES.}
    \label{fig:exc_en_aug_basis}
\end{figure}

\newpage

\section{Statistical summary of the distributions of calculated vertical excitation energies}

\begin{table}[]
    \caption{Statistical summary of the distribution of $n \rightarrow \pi^*$ vertical transition energies (eV) calculated for acetone across the different snapshots, including mean, mode, median and standard error of the mean (SEM). Results are compared for selected active spaces (CASSCF (12,10) and DMRG (24,22)) and solvation models.}
    \label{t:acetone_mean_mode_median_sigma}
    \centering
    \begin{tabular}{|c|c|c|c|c|c|}
    \hline
  Active space & Solvent model & Mean (eV)& Mode (eV)& Median (eV) & SEM (eV)\\
 \hline   
 (12,10) &  FQ$^a$ & 5.30 & 5.31 & 5.24 & 0.02 \\
 (12,10) &  FQ$^b$ & 5.69 & 5.75 & 5.74 & 0.02 \\
 \hline
 (24,22) &  ESPF & 4.96 & 5.01 & 4.96 & 0.01 \\
 (24,22) &  FQ$^a$ & 4.89 & 4.90 & 4.89 & 0.01 \\
 (24,22) &  FQ$^b$ & 5.06 & 4.92 & 5.07 & 0.01\\
 \hline
    \end{tabular}
\end{table}

\begin{table}[]
    \caption{Statistical summary of the calculated $\pi \rightarrow \pi^*$ vertical transition energies (eV) of DCBT across different MD snapshots, including the mean, mode, median, and standard error of the mean (SEM) for the (30,27) active space and the FQ$^b$ solvent model.}
    \label{t:DCBT_mean_mode_median_sigma}
    \centering
    \begin{tabular}{|c|c|c|c|c|c|}
    \hline
  Active space & Solvent model & Mean (eV)& Mode (eV)& Median (eV) & SEM (eV)\\
 \hline   
 (30,27) &  FQ$^b$ & 3.32 & 3.36 & 3.30 & 0.02\\
 \hline
    \end{tabular}
\end{table}

\newpage

\section{Convergence of vertical excitation energies with respect to the number of snapshots extracted from the MD simulation}\label{sec:si-conv}

\begin{table}[]
    \caption{Mean excitation energy of the $n \rightarrow \pi^*$  (eV) of acetone, with standard error and 95\% confidence interval, evaluated over 50, 100, 150, and 200 snapshots.}
    \label{t:acetone_conv}
    \centering
    \begin{tabular}{|c|c|c|c|c|}
    \hline
  Level of theory & Number of snapshots & Mean (eV) & SEM (eV)&  95\% (eV) C.I. \\
   \hline
 CASSCF(12,10)/ FQ$^a$ & 50 & 5.679 & 0.050 & 0.098\\
  & 100 & 5.704 & 0.035 & 0.068\\
  & 150 & 5.711 & 0.027 & 0.054\\
  & 200 & 5.688 & 0.024 & 0.048\\
 \hline
 CASSCF(12,10)/ FQ$^b$ & 50 & 5.254 & 0.043 & 0.085 \\
  & 100 & 5.307 & 0.030 & 0.059 \\
  & 150 & 5.303 & 0.024 & 0.047 \\
  & 200 & 5.302 & 0.020 & 0.040\\
 \hline
 DMRG(24,22)/ESPF & 50 & 4.923 & 0.026 & 0.051\\
  & 100 & 4.953 & 0.017 & 0.033 \\
  & 150 & 4.954 & 0.013 & 0.026 \\
  & 200 & 4.957 & 0.011 & 0.021\\
 \hline
 DMRG(24,22)/FQ$^a$ & 50 & 4.861 & 0.025 & 0.050 \\
  & 100 & 4.892 &  0.018 & 0.035\\
  & 150 & 4.888 & 0.014 & 0.027 \\
  & 200 & 4.890 & 0.011 & 0.022 \\
 \hline
 DMRG(24,22)/FQ$^b$ & 50 & 5.051 & 0.030 & 0.060 \\
  & 100 & 5.074 & 0.021 & 0.041 \\
  & 150 & 5.067 & 0.016 & 0.032 \\
  & 200 & 5.064 & 0.014 & 0.028 \\
 \hline
    \end{tabular}
\end{table}

\begin{table}[]
    \caption{Mean DMRG(30,27)/FQ$^b$ excitation energy of the $\pi \rightarrow \pi^*$ transition (eV) of DCBT, with standard error and 95\% confidence interval, evaluated over 50, 100, 150, and 200 snapshots.}
    \label{t:DCBT_conv}
    \centering
    \begin{tabular}{|c|c|c|c|c|}
    \hline
  Level of theory & Number of snapshots & Mean (eV) & SEM (eV)& 95\% C.I. (eV)  \\
   \hline
 DMRG(30,27)/FQ$^b$ & 50 & 3.374 & 0.036 & 0.071 \\
  & 100 & 3.351 & 0.024 & 0.047 \\
  & 150 & 3.325 & 0.019 & 0.037\\
  & 200 & 3.322 & 0.016 & 0.032\\
 \hline
    \end{tabular}
\end{table}

\newpage
\section{Excitation energies and solvatochromic shifts}

\begin{table}[ht!]
    \caption{Excitation energies ($\varepsilon$) and water-to-vacuum solvatochromic shifts ($\delta = \omega_\mathrm{solv} - \omega_\mathrm{gas}$) of the $n \rightarrow \pi^*$ transition of acetone in aqueous solution, computed at the DMRG(24,22)[300]/aug-cc-pVDZ and CASSCF(12,10)/aug-cc-pVDZ levels of theory using different solvent models (ESPF, FQ$^{(a,b)}$), together with experimental values.}
    \label{t:acetone_exc_en_solv_shift}
    \centering
    \begin{tabular}{|c|c|c|c|}
    \hline
  Active space & Solvent model & $\varepsilon$ (eV) & $\delta$ (eV) \\
 \hline   
 (12,10) &  gas phase & 5.01 & -- \\
 (12,10) &  FQ$^a$ & 5.30 & 0.29 \\
 (12,10) &  FQ$^b$ & 5.69 & 0.68 \\
 \hline
 (24,22) &  gas phase & 4.84 & -- \\
 (24,22) &  ESPF & 4.96 & 0.12 \\
 (24,22) &  FQ$^a$ & 4.89 & 0.05 \\
 (24,22) &  FQ$^b$ & 5.06 & 0.22 \\
 \hline
 Exp. & gas phase & 4.46\cite{renge2009solvent}, 4.48\cite{bayliss1968solvent} & --\\
 Exp. & water & 4.68\cite{renge2009solvent}, 4.69\cite{bayliss1968solvent}& 0.22\cite{renge2009solvent}, 0.21\cite{bayliss1968solvent} \\
 \hline
    \end{tabular}
\end{table}

\begin{table}[ht!]
    \caption{Excitation energies ($\varepsilon$) and solvatochromic shifts ($\delta = \omega_\mathrm{solv} - \omega_\mathrm{gas}$) of the $\pi \rightarrow \pi^*$ transition of DCBT in acetonitrile, computed at the DMRG(30,27)/FQ$^b$[300]/6-31G* level, together with digitized TD-DFT/IEFPCM/def2-TZVP, SA(2)-CASSCF/C-PCM(6,5)/6-31G* literature data and experimental values.}
    \label{t:dcbt_exc_en_solv_shift}
    \centering
    \begin{tabular}{|c|c|c|c|}
    \hline
  Level of theory & Solvent model & $\varepsilon$ (eV) & $\delta$ (eV) \\
 \hline   
 TD-DFT/IEFPCM &  MCH & 2.07 & -- \\
 TD-DFT/IEFPCM &  ACN & 2.45 & -0.38 \\
 \hline
 SA(2)-CASSCF/C-PCM(6,5) & gas-phase & 4.44 & -- \\
 SA(2)-CASSCF/C-PCM(6,5) & DMSO & 3.68 & -0.76 \\
 \hline
 DMRG(30,27)/FQ$^b$ &  gas phase & 3.73 & -- \\
 DMRG(30,27)/FQ$^b$ &  crystal structure & 3.66 & -- \\
 DMRG(30,27)/FQ$^b$ &  ACN & 3.32 & -0.41,-0.34 \\
 \hline
 Exp. & MCH & 2.12 & --\\
 Exp. & ACN & 2.27 & -0.15\\
 \hline
    \end{tabular}
\end{table}

\section{Digitized DCBT absorption spectra from the literature}

Data from references \citenum{paper_DCBT} and \citenum{song_casscf_sa_pcm} were extracted using the WebPlotDigitizer software \cite{WebPlotDigitizer}. In particular, from Figure 2 of ref. \citenum{paper_DCBT}, we selected the left inflection point of each relevant band to reduce the error arising from the neglect of nuclear structure effects in the transition. 

\begin{figure}[ht!]
 \includegraphics[scale=0.35]{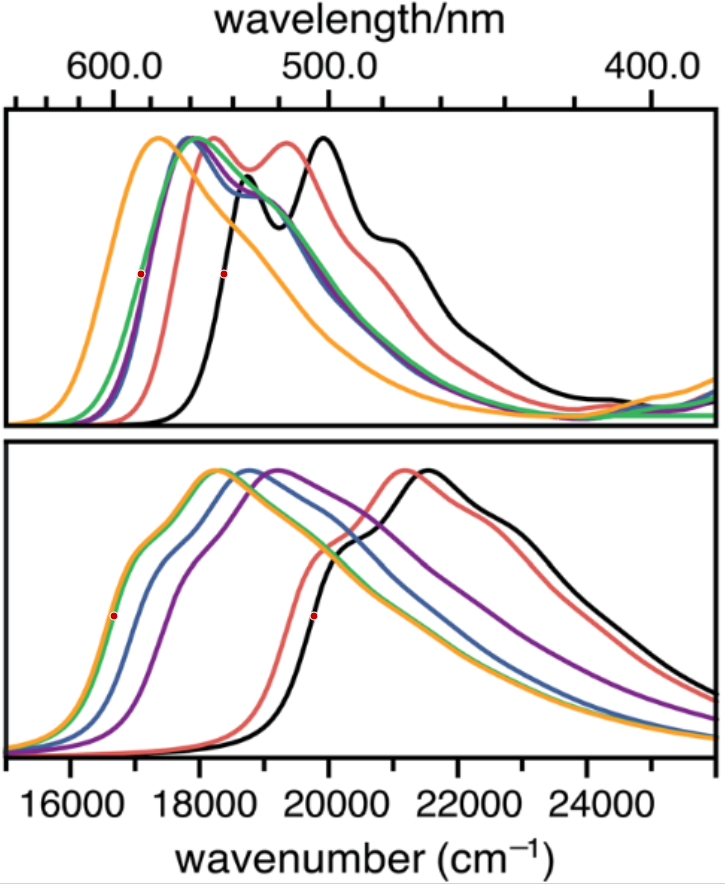} 
    \centering
    \caption{Digitized experimental absorption spectra (top) and calculated absorption spectra at the TDDFT/IEFPCM (def2-TZVP) level (bottom) from fig. 2 of ref. \citenum{paper_DCBT} for DCBT in various solvents. The spectra in green refer to DCBT in acetonitrile (ACN) and the ones in black refer to DCBT in methylcyclohexane (MCH). Red points indicate the left inflection point of each relevant band which are selected to calculate the solvatochromic shift to reduce the error arising from the neglect of nuclear structure effects in the transition.}
    \label{fig:digitized_solv_shift_Mitric}
\end{figure}

\newpage 
\begin{table}[H]
    \caption{Wavenumber ($\tilde{\nu}$) and corresponding frequencies ($\omega$) selected in figure \ref{fig:digitized_solv_shift_Mitric} to calculate the solvatochromic shifts ($\delta$) between acetonitrile (ACN) and methylcyclohexane (MCH) in the $\pi \rightarrow \pi^*$ transition of DCBT. The wavenumber values are extracted using the WebPlotDigitizer software \cite{WebPlotDigitizer} and are converted to frequencies.}
    \label{t:DCBT_exp_and_tddft_shift}
    \centering
    \begin{tabular}{|c|c|c|c|c|c|}
    \hline
   & $\tilde{\nu}_{MCH}(cm^{-1})$ & $\tilde{\nu}_{ACN}(cm^{-1})$ & $\omega_{MCH}(eV)$ & $\omega_{ACN}(eV)$  & $\delta\;(eV)$  \\
   \hline
experimental & 17084 & 18307 & 2.12 & 2.27 & -0.15 \\
TDDFT/IEFPCM (def2-TZVP) & 16666 & 19764 & 2.07 & 2.45 & -0.38 \\
 \hline
    \end{tabular}
\end{table}

\begin{table}[H]
    \caption{Frequencies ($\omega$) selected in fig. 2  of ref. \citenum{song_casscf_sa_pcm} to calculate the solvatochromic shift ($\delta$) between dimethylsulfoxide (DMSO) and gas-phase and the solvatochromic shift ($\Delta$) between toluene (TOL) and gas-phase in the $\pi \rightarrow \pi^*$ transition of DCBT. The frequency values are extracted using the WebPlotDigitizer software \cite{WebPlotDigitizer}.}
    \label{t:DCBT_cas_6_5_shift}
    \centering
    \begin{tabular}{|c|c|c|c|c|c|}
    \hline
    & $\omega_{gas-phase}(eV)$ & $\omega_{DMSO}(eV)$  & $\delta\;(eV)$ & $\omega_{TOL}(eV)$ & $\Delta\;(eV)$ \\
   \hline
SA(2)-CASSCF/C-PCM (6,5)  & 4.44 & 3.68 & -0.76 & 4.04 & -0.40\\
 \hline
    \end{tabular}
\end{table}



\bibliography{biblio}